**Adjustment for Unmeasured Spatial Confounding in Settings of Continuous Exposure Conditional on the Binary Exposure Status: Conditional Generalized Propensity Score-Based Spatial Matching**

Honghyok Kim, Michelle Bell,

School of the Environment, Yale University, New Haven, CT, the United States

Email: honghyok.kim@yale.edu

Telephone: (203) 432-9869

Address: 195 Prospect Street, New Haven, CT, 06511, United States.

**Description of conflict of interest**: We declare there is no conflict of interest.

**Acknowledgement:**

This publication was developed under Assistance Agreement No. RD835871 awarded by the U.S. Environmental Protection Agency to Yale University. It has not been formally reviewed by EPA. The views expressed in this document are solely those of the authors and do not necessarily reflect those of the Agency. EPA does not endorse any products or commercial services mentioned in this publication. Research reported in this publication was also supported by the National Institute On Minority Health And Health Disparities of the National Institutes of Health under Award Number R01MD012769. The content is solely the responsibility of the authors and does not necessarily represent the official views of the National Institutes of Health. This research was supported by Basic Science Research Program through the National Research Foundation of Korea (NRF) funded by the Ministry of Education (2021R1A6A3A14039711)

**ABSTRACT**

Propensity score (PS) matching to estimate causal effects of exposure is biased when unmeasured spatial confounding exists. Some exposures are continuous yet dependent on a binary variable (e.g., level of a contaminant (continuous) within a specified radius from residence (binary)). Further, unmeasured spatial confounding may vary by spatial patterns for both continuous and binary attributes of exposure. We propose a new generalized propensity score (GPS) matching method for such settings, referred to as conditional GPS (CGPS)-based spatial matching (CGPSsm). A motivating example is to investigate the association between proximity to refineries with high petroleum production and refining (PPR) and stroke prevalence in the southeastern United States. CGPSsm matches exposed observational units (e.g., exposed participants) to unexposed units by their spatial proximity and GPS integrated with spatial information. GPS is estimated by separately estimating PS for the binary status (exposed vs. unexposed) and CGPS on the binary status. CGPSsm maintains the salient benefits of PS matching and spatial analysis: straightforward assessments of covariate balance and adjustment for unmeasured spatial confounding. Simulations showed that CGPSsm can adjust for unmeasured spatial confounding. Using our example, we found positive association between PPR and stroke prevalence. Our *R* package, *CGPSspatialmatch,* has been made publicly available.

## INTRODUCTION

Propensity scores (PS) are widely used to estimate causal effects of binary treatments/exposures with observational data[1] and extended for continuous treatments/exposures, referred as generalized PS (GPS)[2,3]. Valid estimation of causal effects using (G)PS relies on the assumption of no unmeasured confounding[4]. Exposures, health outcomes, and unmeasured confounders commonly exhibit spatial patterns, introducing potential unmeasured spatial confounding[5-8]. In regression-based methods, unmeasured spatial confounding may be addressed through adjustment for spatially correlated errors such as spatial random effects[6,7,9,10]. Researchers may follow this practice in estimating (G)PS to minimize unmeasured spatial confounding[11]. PS trimming may also be useful to address unmeasured confounding in general[12].

Researchers are often confronted with settings where continuous exposure is conditional on binary exposure status. To our knowledge, (G)PS methods are limited to settings for either binary or continuous exposure, but not for continuous exposure conditional on its binary exposure status. Researchers may use GPS by treating such exposure as a single continuous variable to account for the degree of exposure in which the binary exposure status is subsumed. However, GPS estimation may be biased because continuous exposure conditional on binary exposure status may have bimodal or skewed exposure distribution. Examples may include accessibility to emergency services (e.g., emergency medical care, fire stations)[13-15], exposure to nearby environmental factors (e.g., power plants, oil and gas extractions, high voltage cables, overhead powerlines, greenness)[16-20], and surrounding built environments (e.g., walkability, recreational facilities, transportation networks)[21]. If exposure focuses only on proximity to exposure, analysis may include a continuous variable indicating distance from the location of an observational unit (e.g., residence of a subject) or a binary variable indicating whether exposure is located within a specified radius (i.e., "buffer") from the unit. Researchers may want to consider proximity along with other quantity-wise characteristics of exposure (e.g., number of ambulance/fire trucks available, air pollution level, number of oil and gas wells nearby, level of greenness nearby, number of recreation facilities nearby). In this case, exposure may be defined as a continuous variable conditional on a binary variable (e.g., buffer). For example, inverse-distance weighted average number of oil production wells within a 10-mile from maternal residence[19] and the sum of production volumes at oil and gas wells within 1-km from maternal residence[18] were used to assess oil production wells in relation to birth outcomes. Other examples are the number of fast-food restaurants within 1-mile from schools in relation to childhood obesity[22], and tree cover within 50m of patient residences in relation to mortality during tuberculosis treatment[23]. Spatial accessibility measured by the two-step floating catchment area method[24], which is a continuous metric where travel time from an observational unit to healthcare providers within a pre-specified maximum buffer distance is a determinant of this metric, was used to explore associations with many health outcomes[13-15]. All these continuous exposure metrics are defined as 0 if the exposure does not exist within a specified buffer from observational units (i.e., binary exposure status). Otherwise, the metrics are defined as >0 with a continuous variable. Thus, exposure distribution can be bimodal and/or skewed (Figure 1). Both binary and continuous exposures can commonly exhibit spatial patterns raising concerns of unmeasured spatial confounding.

To estimate causal effects of such exposure while addressing unmeasured spatial confounding, we introduce a new GPS-matching method. We revamp GPS estimation for continuous exposure conditional on the binary exposure status. GPS is estimated by separately estimating PS and GPS conditional given the binary exposure status and then combining them. We refer to GPS conditional given the binary exposure status as conditional GPS (CGPS). Spatial information is integrated into estimation of PS and CGPS. Exposed observational units are matched to unexposed units by spatial proximity and GPS. We refer to our method as CGPS-based spatial matching (CGPSsm). The benefits of CGPSsm include mimicking randomized clinical trials with covariate balancing and benefits of spatial analyses integrating spatial information in adjusting for unmeasured spatial confounding.

Our method development was motivated by emerging research for which unmeasured spatial confounding may be of concern in evaluating causal effects of oil and gas development and refining[18,19,25], although our method is applicable to other exposures conditional on their binary exposure status. Our motivating example is to investigate possible links between proximity to refineries with high petroleum production and refining (PPR) and stroke risk in the southeastern United States (U.S.). The high stroke risk in this region was first recognized in the 1960s, giving rise to the name "Stroke Belt"[26]. Stroke risk had decreased over time, but the risk is still higher in the Stroke Belt, and the region of higher risk has expanded to include the eastern Texas[27]. We note that this region largely overlaps with the Petroleum Administration of Defense Districts-3 region (PADD3) including Alabama, Arkansas, Mississippi, Louisiana, Texas, and New Mexico where approximately two-thirds of U.S. petroleum production and refining (PPR) takes place. Byproducts in a series of operations (i.e., oil extraction, transportation, refinement, storage and distribution) include air, water, and soil pollution[28,29], which may be linked to cardiovascular and cerebrovascular risks[30,31]. Figure 2 presents locations of 59 operating petroleum refineries for 2015–2017 in the six PADD-3 states and Oklahoma based on data from the U.S. Energy Information Administration (US-EIA). Oklahoma was included due to proximity of their refineries to northern Texas. While (G)PS matching is useful in that it does not strongly depend on a parametric model in confounding adjustment and enables empirically assessed covariate balance of measured confounders[11,32-34], we should be careful about unmeasured spatial confounding in terms of both proximity to refineries and the level of emissions. The entrenched stroke risk in the southeastern U.S. is not fully explained by many factors such as socio-demographics, health behaviors (e.g., smoking, diet, physical activity), hypertension, diabetes, and structural racism[27,35,36]. Figure 3 shows semi-variograms for census tract-level stroke prevalence in the seven states in 2018 and for residual of this prevalence after adjustment for measured census tract-level potential confounders. These plots suggest potential unmeasured spatial confounding if the unmeasured determinants of these spatial patterns are correlated with but not determined by proximity to refineries nor pollutants emitted[7,8]. Standard error regarding the association between PPR and stroke prevalence may also be underestimated[7].

Below we introduce CGPSsm and two applications of this method: a simulation and a real-world case using our motivating example.

## CONDITIONAL GENERALIZED PROPENSITY SCORE-BASED SPATIAL MATCHING

### Notations and Average Treatment Effects in the Treated

Let $Z_i^b$ denote a binary indicator of whether the *i*th of *n* observational unit (e.g., subject, area) is exposed ($Z_i^b$=1) or unexposed ($Z_i^b$=0). Let $Z_i^c$ be a continuous variable of the degree of exposure ($Z_i^c$>0 if $Z_i^b$=1; $Z_i^c$=0 if $Z_i^b$=0). In our motivating example, $Z_i^b$ is the indicator of whether any refinery is located within a specified distance from census tract $i$. $Z_i^c$ is the actual petroleum production that is assumed to be proportionate to emissions of pollutants from refineries in census tract $i$. Let $Y_i$ be a continuous variable of a health outcome (e.g., stroke prevalence) of census tract $i$.

Suppose that each observational unit has potential outcomes by an exposure contrast, $Z_i^c = [0, W]$. We denote $Y_i(w)$ as the potential outcome by $Z_i^c$, $w \in [0, W]$. Assuming that indexing is random, we omit subscript $i$. We define the average treatment effect in the treated (ATT) for $Z^c$ as

$$ATT = E[Y(w) - Y(0)|Z^b = 1]$$

In estimation, ATT is averaged over $w$ and presented as per $\Delta$ unit increase of $Z^c$. ATT answers causal questions such as "to what degree $Z$ increases the risk of $Y$ in the exposed?" by a potential experiment: what if the exposed had not been exposed?

To estimate ATT, the weak unconfoundness assumption is needed[3]. Let $\boldsymbol{C}$ be a minimal set of confounding variables. The weak unconfoundness assumption is:

$$Y(w) \perp Z^c|\boldsymbol{C} \ \ for\ w \in [0, W]$$

Under this assumption, ATT can be estimated by comparing health outcomes across $Z^c$, conditional on $\boldsymbol{C}$. As the dimension of $\boldsymbol{C}$ increases, this assumption becomes more difficult to meet. Researchers compress the information of $\boldsymbol{C}$ into GPS as a balancing score that can be used for confounding adjustment. Let $f(Z^c|\boldsymbol{C})$ denote GPS as the conditional density of $Z^c$ given $\boldsymbol{C}$[3].

With the weak unconfoundness assumption and theorems proved by Hirano and Imbens (2004)[3], we can write ATT using GPS as:

for $w \in (0, W]$, $E_r\left[E\left[Y \middle| Z^c = w, Z_{cf}^b = 1, f(Z_{cf}^c = w|\boldsymbol{C}) \ = r\right] - E\left[Y \middle| Z^c = 0, Z_{cf}^b = 1, f(Z_{cf}^c = w|\boldsymbol{C}) \ = r\right]\right]$ (See Appendix). $Z_{cf}$ is counterfactual $Z$.

$E\left[Y \middle| Z^c = w, Z_{cf}^b = 1, f(Z_{cf}^c = w|\boldsymbol{C}) \ = r\right]$ indicates the potential outcome of the exposed if the exposed had been exposed, which is observed as the outcome of the exposed, $E[Y|Z^c = w]$ and $Z_{cf} = Z = w$. $E\left[Y \middle| Z^c = 0, Z_{cf}^b = 1, f(Z_{cf}^c = w|\boldsymbol{C}) \ = r\right]$ indicates the observed outcome of the unexposed, $E[Y|Z^c = 0]$, serving as the potential outcome of the exposed if the exposed had not been exposed. This potential outcome is unobservable. So, we assume that the observed outcome of the unexposed can serve as the potential outcome, depending on their GPS, $f(Z_{cf}^c = w|\boldsymbol{C})$. This technique is analogous to standard PS matching where the exposed is matched to the unexposed by PS to estimate ATT where the observed outcome of the unexposed serve as the potential outcome of the exposed if the exposed had not been exposed[37]. We introduce estimation of GPS and ATT using CGPSsm below.

**Estimation of ATT**

Let $\boldsymbol{U}$ be a set of observed confounders and $\boldsymbol{U}$ a set of unmeasured confounders that have spatial patterns such that $\boldsymbol{C} = (\boldsymbol{X}, \boldsymbol{U})$. We propose CGPSsm to estimate ATT of $Z^c$ while addressing unmeasured spatial confounding.

***Double-Matching by spatial proximity and GPS***

We augment confounding adjustment by utilizing spatial information of observational units in matching and estimation of GPS. The first step of CGPSsm is to match each exposed unit to unexposed unit(s) by spatial proximity with replacement, referred to as one-to-n distance-matching. Proximity is defined as distance measure, $d$, between two locations. Spatially neighboring areas have similar characteristics such that matching by proximity will have better covariate balance in $\boldsymbol{C} = (\boldsymbol{X}, \boldsymbol{U})$.

CGPSsm further addresses imbalance in $\boldsymbol{C}$ by GPS-matching in each distance-matched stratum. Several matching techniques may be applied. We focus on one-to-one nearest-neighbor matching with/without replacement and one-to-one nearest neighbor caliper matching with/without replacement[38]. Finally, CGPSsm creates a one-to-one distance- and GPS-matched sample with/without replacement. For this, in GPS-matching without replacement, if the same unexposed units exist in multiple distance-matched strata after one-to-n distance-matching with replacement, once an unexposed unit is matched to the exposed unit in one stratum by GPS, then the same unexposed unit in the other strata is deleted. For readers interested in visual illustrations, see eFigure 1.

Matching with replacement may reduce bias more than matching without replacement when unexposed units are sparse[33] but it may reduce precision if fewer unexposed units in the original sample are matched (e.g., multiple exposed units are matched to one unexposed unit)[33].

***Estimation of GPS***

GPS estimation may be biased if the skewed/bimodal distribution of $Z^c$ due to $Z^b$ is not adequately considered. In CGPSsm, GPS is estimated by separately estimating PS for $Z^b$ and estimating CGPS that is GPS for $Z^c$ given $Z^b$. By the law of total probability, GPS can be decomposed as:

$$f(Z^c = w|\boldsymbol{C}) = P[Z^b = 1|\boldsymbol{C}] \times f(Z^c = w|\boldsymbol{C}, Z^b = 1) + P[Z^b = 0|\boldsymbol{C}] \times f(Z^c = w|\boldsymbol{C}, Z^b = 0)$$

We are interested in only $w > 0$. When $w > 0$, $f(Z^c = w|\boldsymbol{C}, Z^b = 0) = 0$, such that

$$f(Z^c = w|\boldsymbol{C}) = P[Z^b = 1|\boldsymbol{C}] \times f(Z^c = w|\boldsymbol{C}, Z^b = 1)$$

We introduce estimation of CGPS. Of the exposed, CGPS can be estimated as standard GPS estimation[3,39]. We assume that CGPS follows a normal distribution and,

1. Fit a CGPS model to predict $Z^c$ and $\boldsymbol{C}$ for *only the exposed* and get $\hat{\boldsymbol{\beta}}$ and $\hat{\sigma}$.
2. CGPS of the exposed is $f\left(Z^c_{cf} = w \middle| \boldsymbol{C}, Z^b_{cf} = 1\right) = \frac{1}{\sqrt{2\pi\hat{\sigma}^2}} \exp\left(-\frac{1}{2\hat{\sigma}^2}\left(w - (\hat{\boldsymbol{\beta}}\boldsymbol{C}_{\boldsymbol{Z^b=1}})\right)^2\right)$

where $\hat{\boldsymbol{\beta}}$ is a set of regression coefficients to predict $Z^c$, $\hat{\sigma}$ is a standard deviation of residuals, and $\boldsymbol{C}_{\boldsymbol{Z^b=1}}$ is $\boldsymbol{C}$ of the exposed.

We estimate CGPS of the unexposed by

$$f\left(Z^c_{cf} = w \middle| \boldsymbol{C}, Z^b_{cf} = 1\right) = \frac{1}{\sqrt{2\pi\hat{\sigma}^2}} \exp\left(-\frac{1}{2\hat{\sigma}^2}\left(w - \hat{\boldsymbol{\beta}}\boldsymbol{C}_{\boldsymbol{Z^b=0}}\right)^2\right)$$

where $\boldsymbol{C}_{\boldsymbol{Z^b=0}}$ is $\boldsymbol{C}$ of the unexposed. In the original dataset, there are many different values of $w$ over the exposed. After distance-matching, there would be many strata for the distance-matched pairs of one exposed unit to one or several unexposed unit(s). In each stratum, CGPS of the unexposed is estimated according to $w$ of the exposed. GPS is then estimated. eFigure 1 illustrates this process.

To build PS and CGPS models, spatial coordinates may be used to adjust for $U$ such as spatial regression models and gradient boosting algorithms[11]. This augments spatial confounding adjustment in addition to distance-matching. We consider generalized additive models (GAM) with spatial smoother using *mgcv* R package[40] and the extreme gradient boosting (XGBoost) using *xgboost* R package[41] including coordinates as covariates. Spatial models with a random effect as implemented using *spBayes* R package[42] may be an alternative but we did not consider this because of its poor performance in PS matching to adjust for spatial confounding[11].

***Diagnostics of Covariate Balance***

Within each stratum of the matched pairs, one exposed unit is compared to one unexposed unit to estimate ATT. Therefore, we suggest standardized mean difference (SMD) of $\boldsymbol{X}$ between the exposed and unexposed[34,43]. Cut-off values to check covariate balance of 0.1 or 0.25 are often used in PS methods[43]. However, there is no clear threshold although use of a smaller cut-off may intuitively result in stronger adjustment[32]. Correlation in the original dataset with causal knowledge[8] may be informative to build a CGPS model and thereby to achieve covariate balance regarding SMD[44,45].

***Selecting spatial distance threshold in the distance-matching and caliper in the GPS-matching***

Researchers can pick $d$ to augment confounding adjustment to address unmeasured spatial confounding. The lower $d$ matches exposed units to more closely located unexposed units and potentially makes stronger adjustment for $U$. Visual inspection of semi-variograms (Figure 3) can inform selection of $d$. In our example, we use $d$=0.1. Users may select the lowest $d$ that minimizes SMD across different $d$ values to address covariate balance in $\boldsymbol{X}$ as well.

Too low $d$ may result in many exposed units being unmatched. The omission of unmatched exposed units in analysis can lead to loss of precision[33,46]. This can also alter the estimand if ATT for the dropped exposed units is not equal to ATT for the included exposed units[33].

In caliper matching, researchers select caliper width. We refer to $cw$ as a factor of the standard deviation of GPS. The selection of $cw$ is important for covariate balancing. Intuitively, a tighter caliper may reduce bias and precision[33]. A too narrow caliper with low $d$ may result in many exposed units being unmatched. Again, the omission of unmatched exposed units in analysis can also alter the estimand[33]. For $cw \longrightarrow \infty$, nearest neighbor caliper matching becomes nearest neighbor matching.

### *R Package*

CGPSsm is provided as *R* package, *CGPSspatialmatch* in the first author's GitHub (https://github.com/HonghyokKim/CGPSspatialmatch). An illustrative example with R codes is provided at https://hkimresearch.com.

## SIMULATION STUDY

### Methods

We conducted a simulation study to evaluate CGPSsm.

### *Data generation*

We generated three i.i.d confounders ($X_1, X_2, X_3$) from the standardized normal distribution and one spatial confounder ($U$) from a Gaussian process with the Matérn covariance function. Nine pairs of the two spatial parameters ($k$=smoothness, $\pi$=range) in the Matérn covariance function were considered: $k$=0.1, 0.5, 1; $\pi$=0.1, 0.5, or 1. We call these the nine simulation scenarios. Higher $k$ reflects smoother spatial patterns. Higher $\pi$ reflects more correlated spatial units. See Minasny and McBratney (2005)[47] for details. eFigure 2 shows nine spatial confounding patterns. For each spatial pattern, we generated 200 samples with 484 (22×22) fixed locations. $Z^b$ was generated using logistic regression with the four confounders (i.e., approximately 15% exposed). $Z^c$>0 was generated using linear regression with the four confounders if $Z^b$=1. Otherwise, $Z^c$=0. $Y$ was generated using Poisson regression with the four confounders and $Z^c$. More details are in eAppendix 1.

### *CGPSsm*

We conducted CGPSsm with $U$ measured, referred to as true CGPSsm, and CGPSsm with $U$ unmeasured. One-to-n distance-matching with replacement was conducted with $d$ as standardized Euclidian distance of 0.1. For GPS-matching, one-to-one nearest neighbor matching with/without replacement and one-to-one nearest neighbor caliper matching with/without replacement were conducted. We fit PS models using either a GAM with a binomial distribution and the logistic link including $X_1$, $X_2$, $X_3$, and a spatial smoother or XGBoost including $X_1$, $X_2$, $X_3$, and coordinates as covariates. We used grid search with cross-validation to select hyperparameters of XGBoost, which is embedded in our *R* package. We fit CGPS models using GAM with a normal distribution and the identity link including $X_1$, $X_2$, $X_3$ and a spatial smoother using only the exposed of each simulated sample. We generated 500 bootstrapped samples from the distance- and GPS-matched samples. Coefficient estimates and standard error estimates for ATT over the bootstrapped samples were obtained. More details are in eAppendix 1.

### *Other methods*

To illustrate the degree of unmeasured spatial confounding in simulation samples,
we fit Poisson regression including only $X_1$, $X_2$, and $X_3$ as covariates (i.e., no adjustment for spatial confounding). We conducted standard GPS-based inverse probability weighting (IPW) to see the degree of unmeasured spatial confounding when dependency of $Z^c$ on $Z^b$ is not considered. GPS was estimated by a GAM with spatial smoother (Naïve IPW-GAM) or XGBoost (Naïve IPW-XGBoost). GPS was estimated without consideration of dependency of $Z^c$ on $Z^b$. We acknowledge that IPW and conventional Poisson regression is not designed to estimate ATT. The average treatment effect in the population and conditional average effect were set to be identical to ATT in our simulation.

## Results

Bias by unmeasured spatial confounding in our simulation settings was approximately -60% based on Poisson regression without spatial adjustment for all nine simulation scenarios (eFigure 3). For Naïve IPW methods, bias ranged from -20% to -45% in scenarios of $k$=0.5 or 1 with $\pi$=0.5 or 1 but bias increased for other scenarios. Nominal coverage of 95% confidence intervals ranged from 1% to 10.5% for all the scenarios (eFigure 4).

Figure 4 presents the percentage of exposure units matched to unexposed units by CGPSsm methods with replacement. The number in brackets in Figure 4 indicates $cw$ in one-to-one nearest neighbor caliper matching. $cw = \infty$ indicates one-to-one nearest neighbor matching. Fewer exposed units were matched when the caliper was narrower.

Figure 5 presents bias by CGPSsm methods with replacement. Figure 6 presents balance for $X_1$, $X_2$, $X_3$, and $U$ by CGPSsm methods with replacement regarding three selected scenarios. CGPSsm with $U$ known showing bias around 0% when $cw$=0.4, 0.6, or 0.8 confirms that CGPSsm can estimate ATT. The bias was around -5% to -10% when $cw$=∞. Covariates were more balanced with lower $cw$. Note that for PS matching in general, use of the true PS model may not completely reduce bias, depending on matching techniques[38,48].

When $U$ is missing, CGPSsm with GAM and XGBoost showed good adjustment for $U$ in scenarios of smoother spatial patterns of $U$ (i.e., $k$=0.5 or 1) (Figure 5). Specifically, covariates were more balanced with lower $cw$ (Figure 6), resulting in stronger adjustment (Figure 5). Bias ranged roughly from -60% to 5% for caliper matching. For $k$=0.1 where $U$ acted more like a randomly distributed confounder (but is still a spatial confounder), CGPSsm adjusted for $U$ but to a limited degree. For XGBoost, we needed to use higher value of $cw$ than that for CGPSsm with GAM because XGBoost in PS estimation greatly distinguished PS of the exposed and of the unexposed.

Figure 7 presents root mean squared error (RMSE) by CGPSsm methods with replacement. In scenarios of $k$=0.5 or 1 with $\pi$=0.5 or 1, for $cw$ from 0.2 to 0.8, CGPSsm with GAM showed slightly higher RMSE than CGPSsm with *U* known. CGPSsm with GAM generally outperformed CGPSsm with XGBoost.

Figure 8 presents nominal coverage of 95% confidence intervals by CGPSsm methods with replacement. For less biased methods and scenarios (Figure 5), coverage was higher than 95%.

eFigures 5–9 shows CGPSsm methods without replacement regarding bias, RMSE, nominal coverage, covariate balance in selected scenarios, and the percentage of exposure units being matched to unexposed units. They showed good adjustment for *U* while CGPSsm with replacement performed slightly better regarding bias and RMSE.

## APPLICATION STUDY

### Methods

We applied CGPSsm to our motivating example of refineries in relation to stroke prevalence with PPR of refineries conditional on whether refineries are located within a buffer distance from census tracts. A study population is a total of 10381 census tracts in Alabama, Arkansas, Mississippi, Louisiana, Texas, and New Mexico and Oklahoma.

#### *Exposure classification*

Higher overall production capacity is assumed in this example to reflect higher emission of pollutants[18]. To estimate residential exposure to PPR, we considered potential petroleum production of each refinery, as a dimension of exposure in addition to distance between census tracts and refineries. We calculated inverse-distance weighted average of the amount of petroleum production (APP) within a pre-specified buffer distance from the centroid of census tracts as:

$$X_{ct} = \frac{\sum_{r \in R} APP_r \times \frac{1}{D_{ct,r}^2}}{\sum_{r \in R} \frac{1}{D_{ct,r}^2}}$$

$X_{ct}$ is the inverse squared distance-weighted average of APP (barrels/day) at census tract *ct* for 2015-2017. APP was obtained from US-EIA. $D_{c,r}$ is the distance between the centroid of census tract *ct* and refinery *r*. R is a set of refineries within a pre-specified buffer distance from the census tracts' centroid. We calculated $X_{ct}$ values corresponding to 5-km buffers. This threshold was chosen based on air pollution dispersion from smokestacks of refineries and empirical correlation between petroleum production and measured levels of $SO_2$, which is a byproduct of petroleum production and refining[49].

***Outcome and Confounding variables***

Outcome of interest is census tract-level stroke prevalence (%) in 2018. Census tract-level potential confounders include subpopulation distributions by age, sex, race/ethnicity, education, health insurance, median household income, and current smoking. Data sources and details are in eTable 1.

***Statistical analysis***

To estimate census tract-level cross-sectional association of PPR with stroke prevalence in seven states (Figure 1), we used CGPSsm with GAM and CGPSsm with XGBoost. One-to-one nearest neighbor matching with/without replacement were used. One-to-one nearest neighbor with caliper width of 0.2 with/without replacement was used for only CGPSsm with GAM. We also fit a regression model without spatial adjustment to examine degree of unmeasured spatial confounding. More details are in eAppendix 2.

**Results**

Figure 9 presents exposure distribution ($X_{ct}$) that is bimodal: Figure 9A for all census tracts; Figure 9B for exposed census tracts.

Table 1 shows association between PPR exposure and stroke prevalence. Without spatial adjustment, PPR exposure was associated with 0.02 percent point increase (95% CI: -0.03, 0.07) in stroke prevalence per standard deviation increase of PPR. CGPSsm methods consistently showed approximately 0.20 percent point increase in stroke prevalence. The CGPSsm method with the most balanced covariates produced the estimate as 0.20 (0.04, 0.35) while 34.2% of the exposed were dropped in analysis. SMDs for 14 potential confounders (eTable 1) before and after CGPSsm are presented in eFigure 10. CGPSsm with GAM with replacement with $cw$=0.2 or ∞ showed the most balanced covariates with cut-off of ±0.1. Other CGPSsm methods achieved covariate balance with cut-off of ±0.25.

## DISCUSSION

We developed a new GPS-matching method, CGPSsm, to estimate ATT of continuous exposure conditional on binary exposure status while adjusting for unmeasured spatial confounding. CGPSsm enables researchers to empirically assess balance in measured covariates and adjust for unmeasured spatial confounding by leveraging spatially indexed data in matching and GPS estimation procedures. Simulations under various spatial patterns confirm that CGPSsm can greatly reduce unmeasured spatial confounding. We applied CGPSsm to our motivating example, finding that PPR is associated with stroke prevalence in seven southern U.S. states. CGPSsm has potential for many epidemiological studies where exposure has both binary and continuous attributes (e.g., examples in the first section).

We note that many areas and possibilities of CGPSsm remain unaddressed, which warrant future investigations. There exist many other matching techniques (e.g., optimal matching)[38]. We focused on one-to-one nearest neighbor matching with/without caliper and with/without replacement because they are frequently used in epidemiological studies[33,38]. Bootstrapping methods could be studied regarding standard error estimation[50-53]. Theoretical and simulation studies suggest that m-out-of-n bootstrapping may be superior regarding PS matching with replacement[52,53]. Regression techniques to build a PS and CGPS model other than GAM and XGBoost may be applied. PS trimming may be integrated into CGPSsm to augment adjustment for unmeasured confounding[12]. CGPSsm currently shares limitations of PS such as ambiguity of variable specification and selection[44,45,48] and inability to check balance in unmeasured variables[54].

## Appendix

### *Derivations of Average treatment effect in the treated (ATT).*

We note two theorems proved by Hirano and Imbens (2004)[3] using our notations:

Theorem 1. $E[Y(w)] = E[\beta(w, f(Z^c = w|\boldsymbol{C}))]$ and;

Theorem 2. $\beta(w, f(Z^c = w|\boldsymbol{C})) = E[Y(w)|f(Z^c = w|\boldsymbol{C}) = r] = E[Y|Z^c = w, f(Z^c = w|\boldsymbol{C}) = r]$

With the weak unconfoundness assumption and these theorems, we can write ATT using GPS as follows.

$ATT = E[Y(w) - Y(0)|Z^b = 1]$

$= E[Y(w)|Z^b = 1] - E[Y(0)|Z^b = 1]$

$= E_r\left[E[Y(w)|Z^b = 1, f(Z^c = w|\boldsymbol{C}) = r]\right] - E_r\left[E[Y(0)|Z^b = 1, f(Z^c = 0|\boldsymbol{C}) = r]\right]$

$= E_r\left[E[Y(w)|Z^b = 1, f(Z^c = w|\boldsymbol{C}) = r] - E[Y(0)|Z^b = 1, f(Z^c = 0|\boldsymbol{C}) = r]\right]$

$= E_r\left[E[Y|Z^c = w, Z^b = 1, f(Z^c = w|\boldsymbol{C}) = r] - E[Y|Z^c = 0, Z^b = 1, f(Z^c = 0|\boldsymbol{C}) = r]\right]$ (Eq.1)

We leverage a concept in PS matching literature to estimate ATT. $P(Z^b = 1|\boldsymbol{C})$ is used for matching. For PS matching, ATT can be expressed using $P(Z^b = 1|\boldsymbol{C})$ under unconfoundness assumption[37] as $E[Y(1) - Y(0)|Z^b = 1] = E[Y(1)|Z^b = 1] - E[Y(0)|Z^b = 1] = E[Y|Z^b = 1, P(Z^b = 1|\boldsymbol{C})] - E[Y|Z^b = 0, Z^b = 1, P(Z^b = 1|\boldsymbol{C})]$. $E[Y|Z^b = 0, Z^b = 1, P(Z^b = 1|\boldsymbol{C})]$ indicates the observed outcome of the unexposed serving as the potential outcome of the exposed if the exposed had not been exposed. To clarify the notation, we replace $E[Y|Z^b = 0, Z^b = 1, P(Z^b = 1|\boldsymbol{C})]$ with $E[Y|Z^b = 0, Z^b_{cf} = 1, P(Z^b_{cf} = 1|\boldsymbol{C})]$ by introducing $Z_{cf}$, which is a counterfactual *Z*.

Similarly, we replace $E[Y|Z^c = 0, Z^b = 1, f(Z^c = 0|\boldsymbol{C}) = r]$ in Eq.1 with $E[Y|Z^c = 0, Z^b_{cf} = 1, f(Z^c_{cf} = w|\boldsymbol{C}) = r]$. We also use $Z_{cf}$ for the exposed units for clarity. $Z_{cf}$ of the exposed units equals to *Z* because the potential outcome of the exposed if the exposed had been exposed was observed. Then, Eq.1 will be replaced with $E_r\left[E[Y|Z^c = w, Z^b_{cf} = 1, f(Z^c_{cf} = w|\boldsymbol{C}) = r] - E[Y|Z^c = 0, Z^b_{cf} = 1, f(Z^c_{cf} = w|\boldsymbol{C}) = r]\right]$ (Eq.2).

Eq.2 is shown in the main text. Eq. 2 is ATT as follows:

$$E_r\left[E\left[Y\middle|Z^c=w, Z_{cf}^b=1, f\left(Z_{cf}^c=w\middle|\boldsymbol{C}\right)=r\right]-E\left[Y\middle|Z^c=0, Z_{cf}^b=1, f\left(Z_{cf}^c=w\middle|\boldsymbol{C}\right)=r\right]\right].$$

$$=E_r\left[E\left[Y(w)\middle|Z_{cf}^b=1, f\left(Z_{cf}^c=w\middle|\boldsymbol{C}\right)=r\right]-E\left[Y(0)\middle|Z_{cf}^b=1, f\left(Z_{cf}^c=w\middle|\boldsymbol{C}\right)=r\right]\right]$$

$$=E_r\left[E\left[Y(w)\middle|Z_{cf}^b=1, f\left(Z_{cf}^c=w\middle|\boldsymbol{C}\right)=r\right]\right]-E_r\left[E\left[Y(0)\middle|Z_{cf}^b=1, f\left(Z_{cf}^c=w\middle|\boldsymbol{C}\right)=r\right]\right]$$

$$=E\left[Y(w)\middle|Z_{cf}^b=1\right]-E\left[Y(0)\middle|Z_{cf}^b=1\right]=E\left[Y(w)-Y(0)\middle|Z_{cf}^b=1\right]=ATT$$

**Table 1. Associations between exposure to petroleum production refining and stroke prevalence by different methods**

| Methods | Coefficient estimate (95% confidence interval) | Average of absolute SMD for 14 potential confounder variables | Exposed units matched/total exposed units (Dropped %) |
|---|---|---|---|
| Linear regression without spatial adjustment | 0.02 (-0.03, 0.07) | - | - |
| CGPSsm | | | |
| Unmatched sample | - | 0.45 | - |
| GAM (∞) without replacement | 0.22 (0.09, 0.35) | 0.11 | 309/316 (2.2%) |
| GAM (∞) with replacement | 0.19 (0.07, 0.30) | 0.07 | 309/316 (2.2%) |
| GAM (0.2) without replacement | 0.21 (0.01, 0.42) | 0.11 | 159/316 (49.7%) |
| GAM (0.2) with replacement | 0.20 (0.04, 0.35) | 0.06 | 208/316 (34.2%) |
| XGBoost (∞) without replacement | 0.24 (0.11, 0.38) | 0.11 | 309/316 (2.2%) |

Note: The number in brackets indicate $cw$ in one-to-one nearest neighbor caliper matching; $cw = \infty$ indicates one-to-one nearest neighbor matching; SMD=Standardized Mean Difference; SMD for 14 potential confounder variables are presented in eFigure 10.

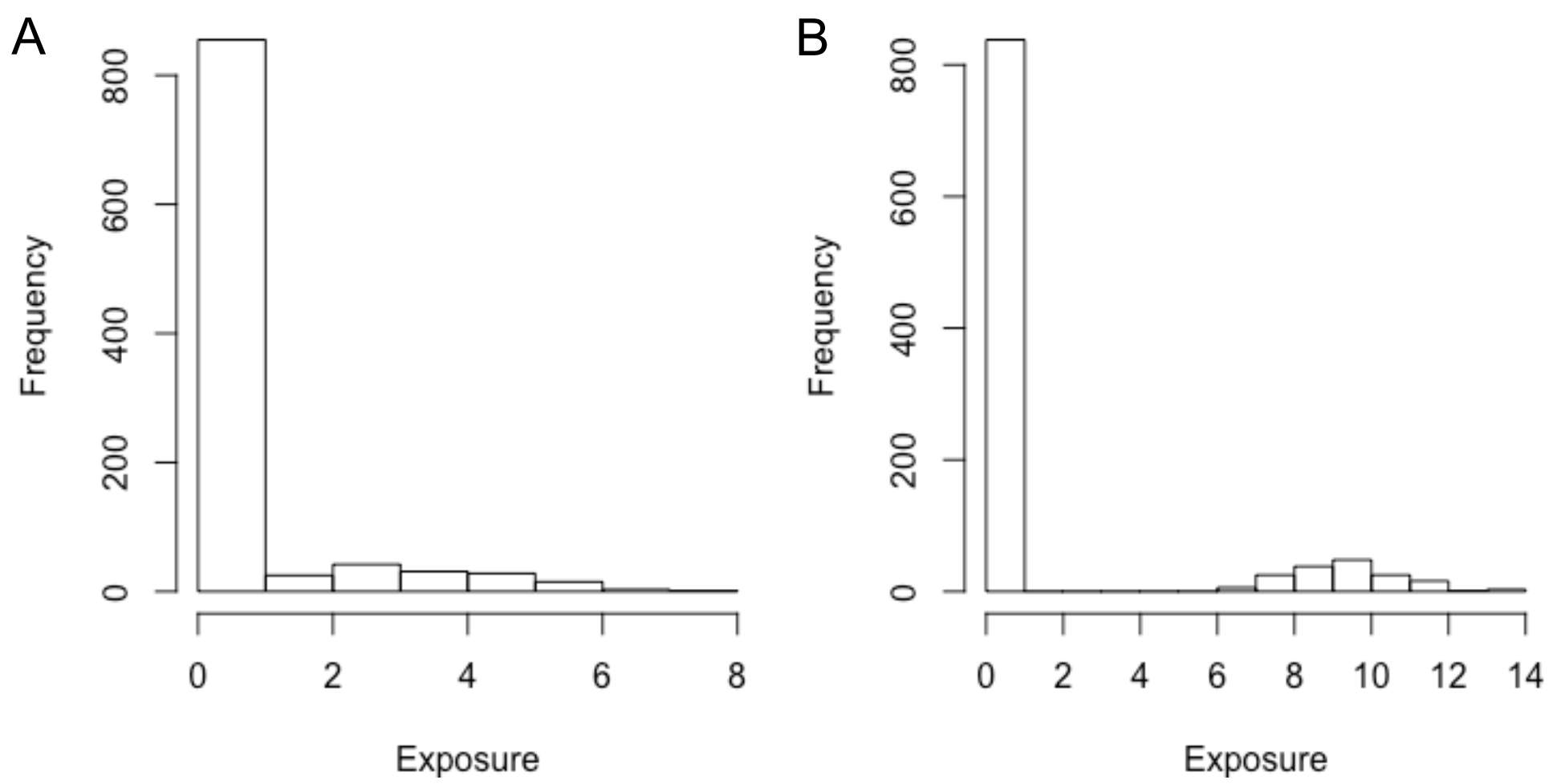

**Figure 1. Hypothetical examples of bimodal and/or skewed distribution of continuous exposure conditional on its binary exposure status**

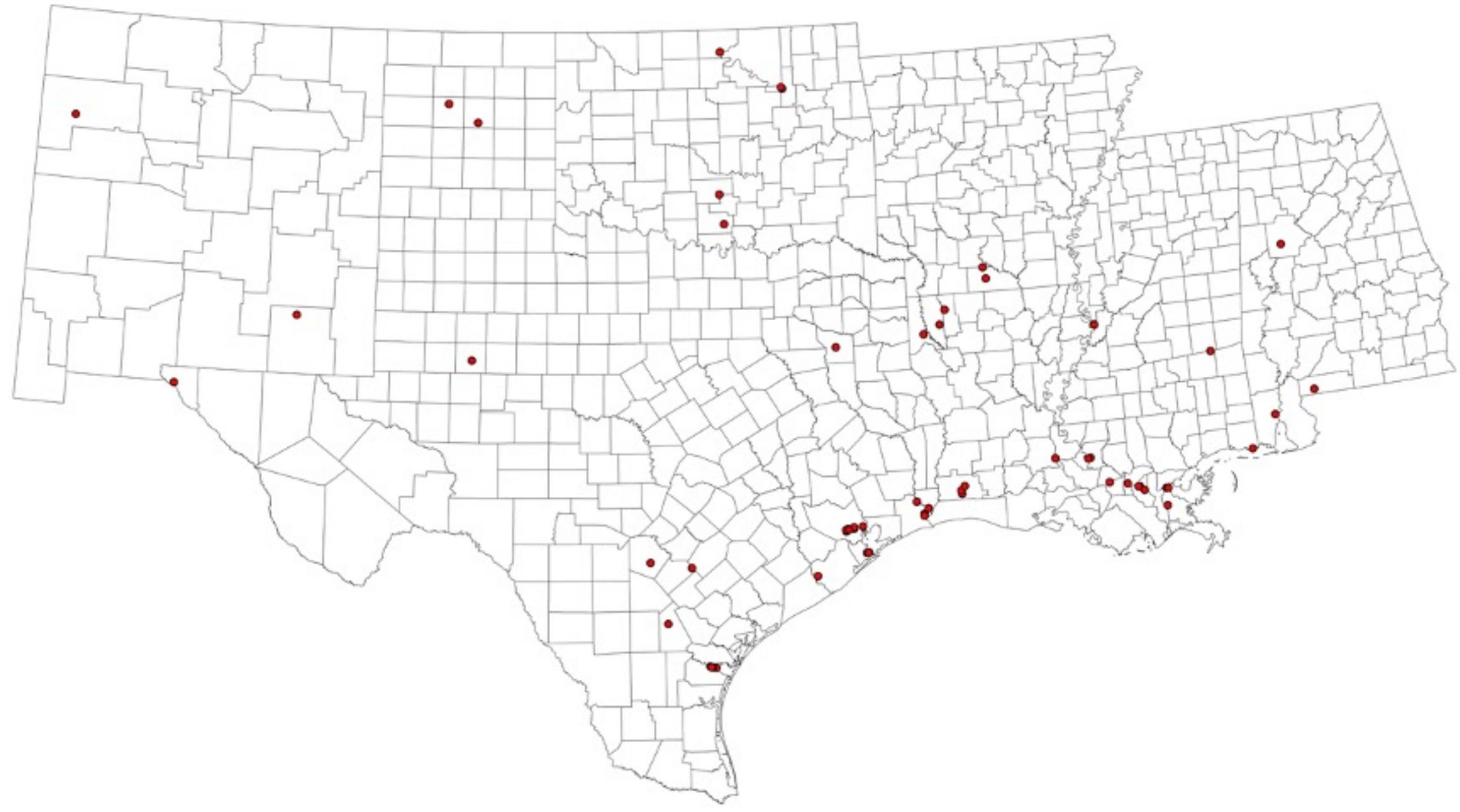

**Figure 2. Locations of petroleum refineries in the Petroleum Administration of Defense Districts-3 region (PADD3) including Alabama, Arkansas, Mississippi, Louisiana, Texas, and New Mexico and Oklahoma.**

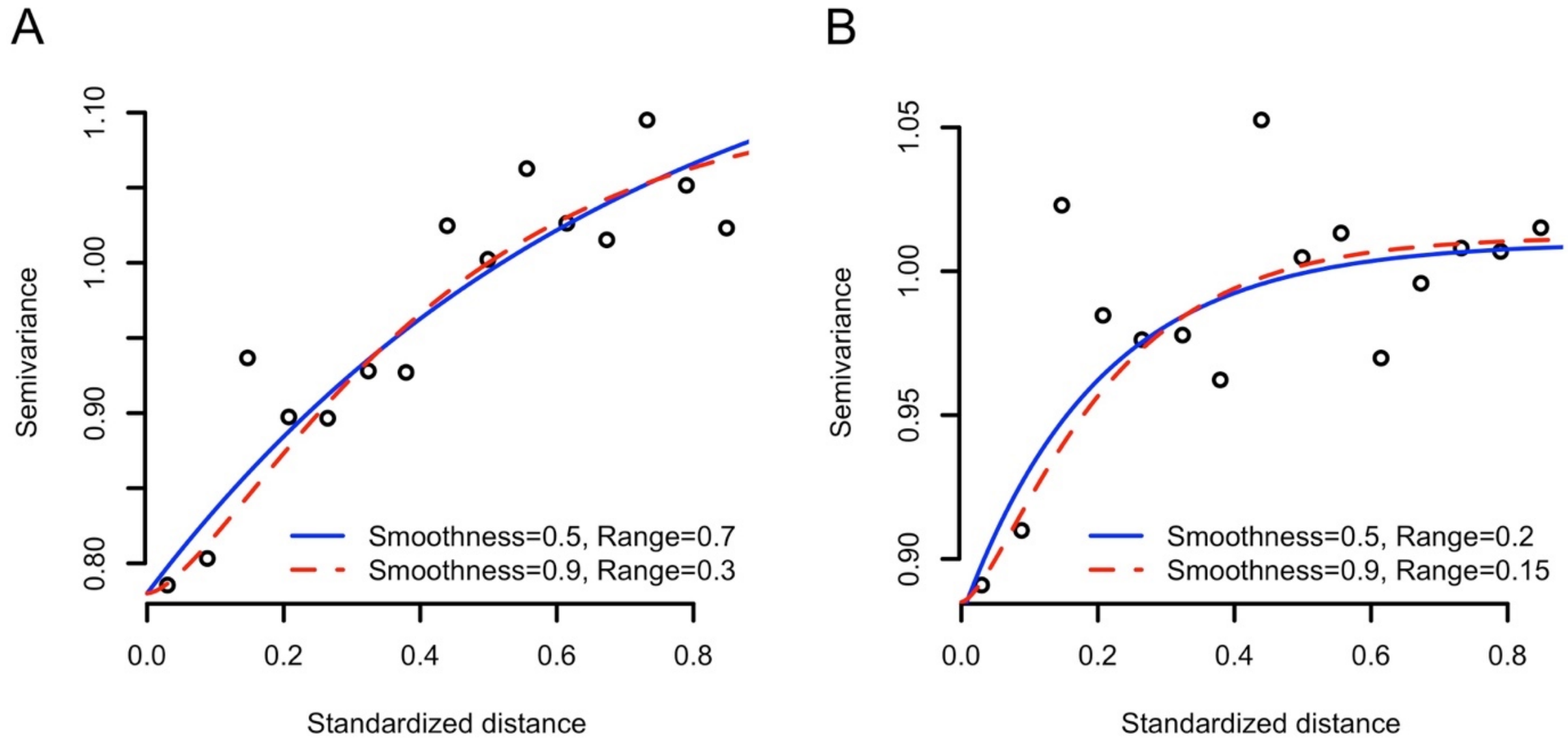

**Figure 3. Variograms for census tract-level stroke prevalence (A) and residual (B) in seven states.**

Note: Residual of stroke prevalence was obtained from a regression model that predicts stroke prevalence using demographic variables, socioeconomic variables, and smoking variable. Data sources and variables are presented in eTable 1. The Matérn covariance function with the smoothness and range parameters are presented.

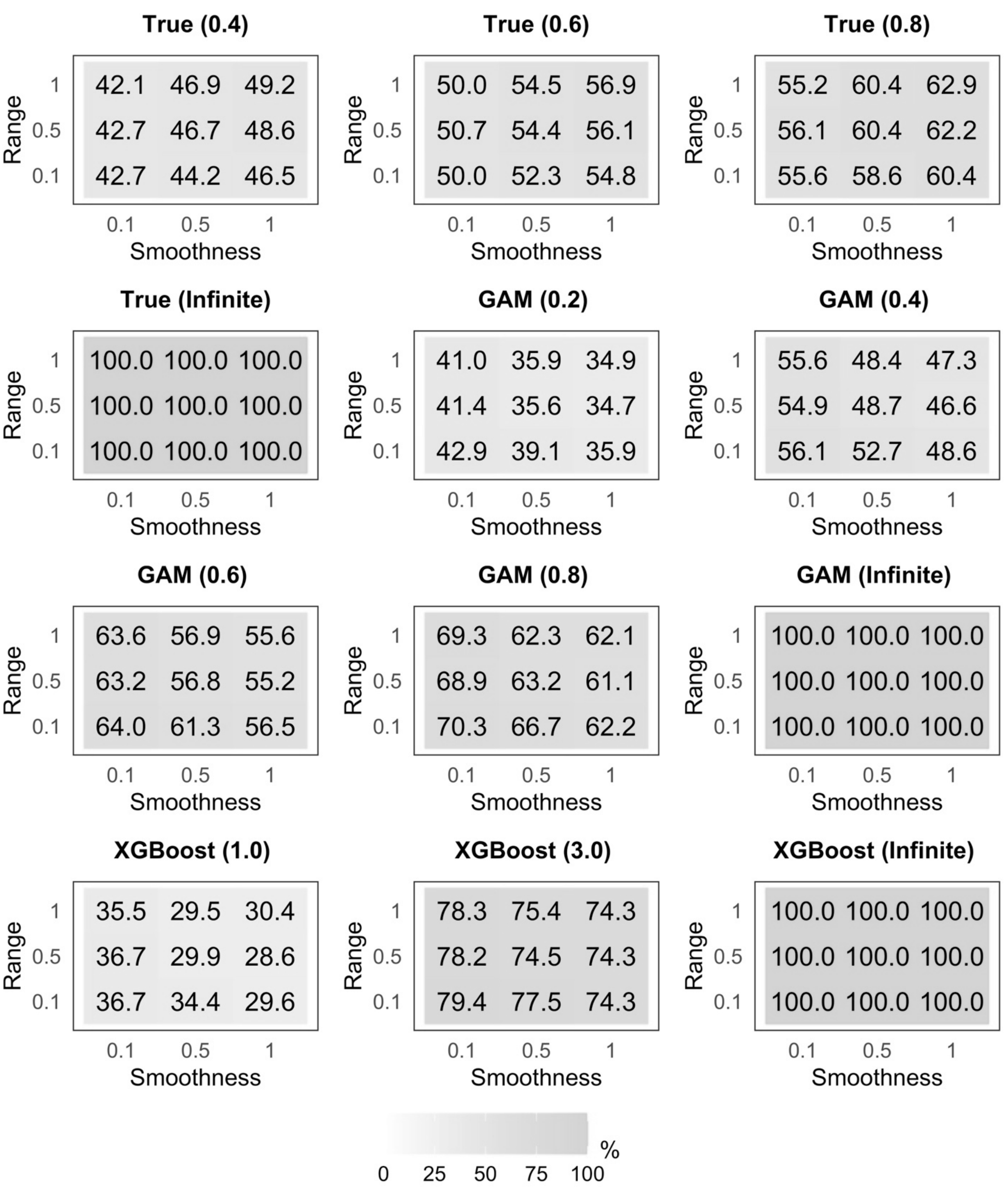

**Figure 4. Percentage of exposed units matched to unexposed units by CGPSsm methods with replacement.**

Note: True=True PS and true CGPS models in CGPSsm (i.e., $U$ is known); GAM=GAM with spatial smoother for PS and CGPS estimations; XGBoost=XGBoost for PS estimation and GAM with spatial smoother for CGPS estimation; The number in brackets indicate $cw$ in one-to-one nearest neighbor caliper matching; $cw = \infty$ indicates one-to-one nearest neighbor matching.

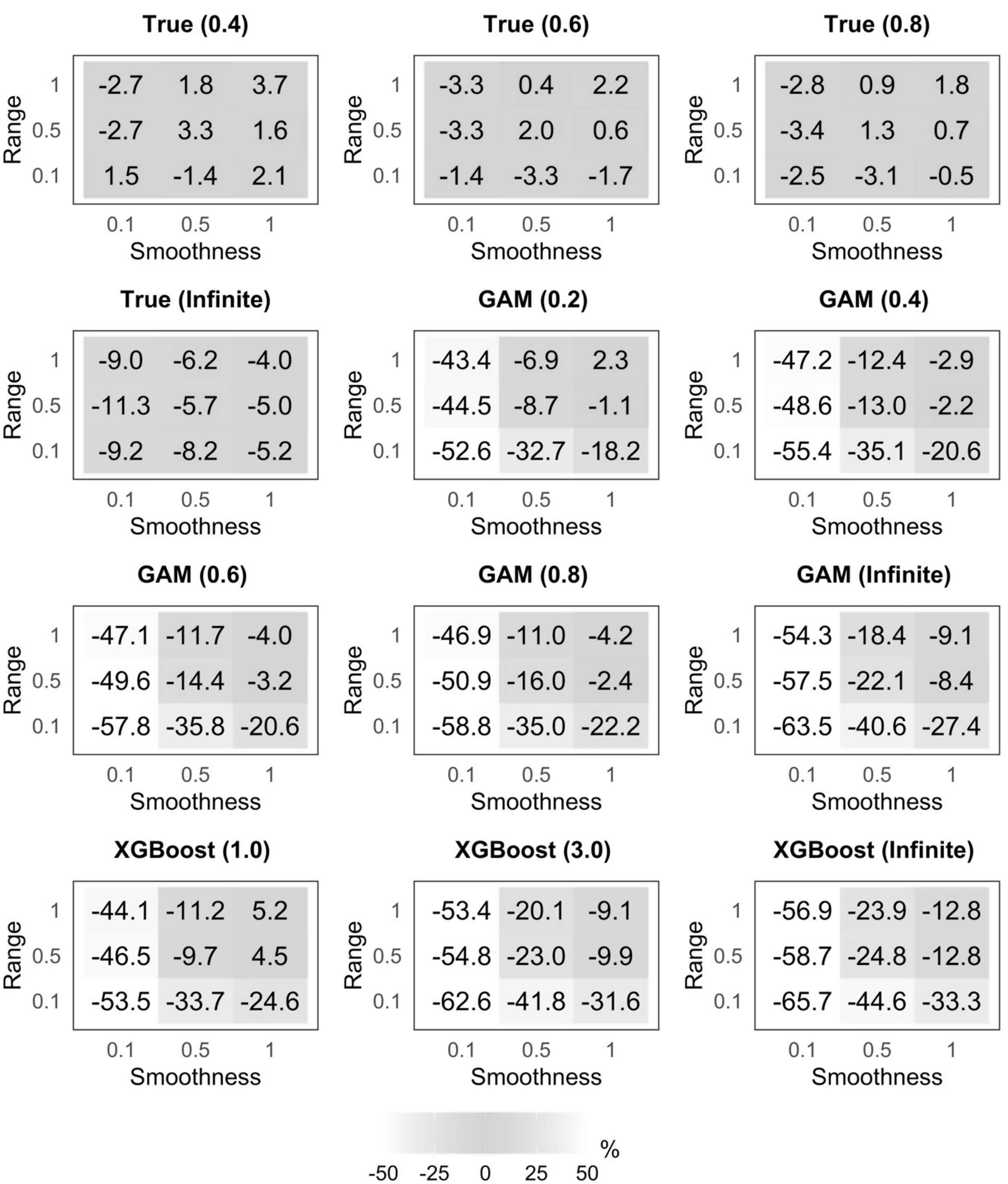

**Figure 5. Bias by CGPSsm methods with replacement.**

Note: True=True PS and true CGPS models in CGPSsm (i.e., *U* is known); GAM=GAM with spatial smoother for PS and CGPS estimations; XGBoost=XGBoost for PS estimation and GAM

with spatial smoother for CGPS estimation; The number in brackets indicate $cw$ in one-to-one nearest neighbor caliper matching; $cw = \infty$ indicates one-to-one nearest neighbor matching.

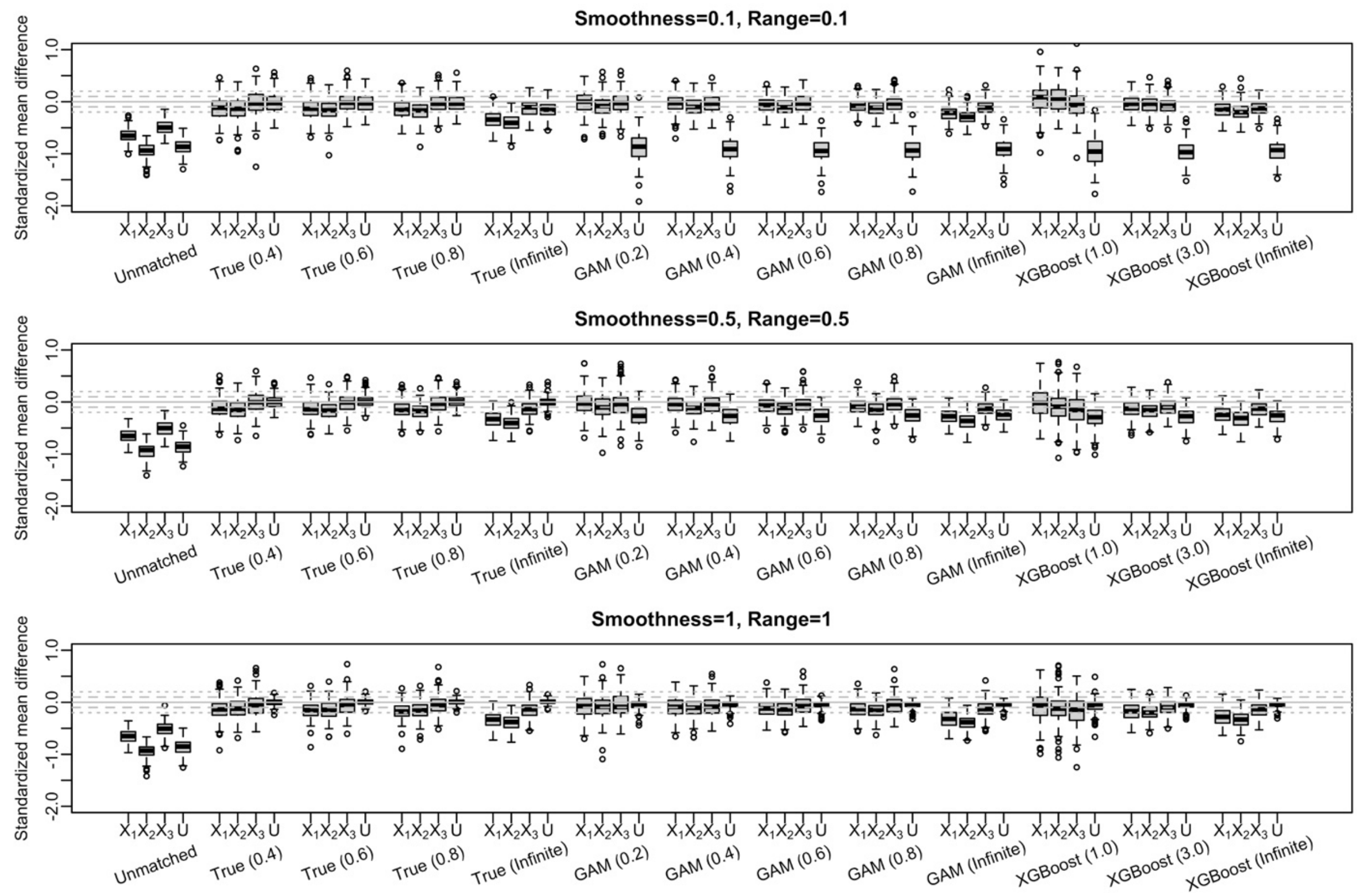

**Figure 6. Covariate balance before and after CGPSsm with replacement.**

Note: True=True PS and true CGPS models in CGPSsm (i.e., *U* is known); GAM=GAM with spatial smoother for PS and CGPS estimations; XGBoost=XGBoost for PS estimation and GAM with spatial smoother for CGPS estimation; The number in brackets indicate $cw$ in one-to-one nearest neighbor caliper matching; $cw = \infty$ indicates one-to-one nearest neighbor matching; Grey dashed lines indicate ±0.1. Grey dotted lines indicate ±0.25.

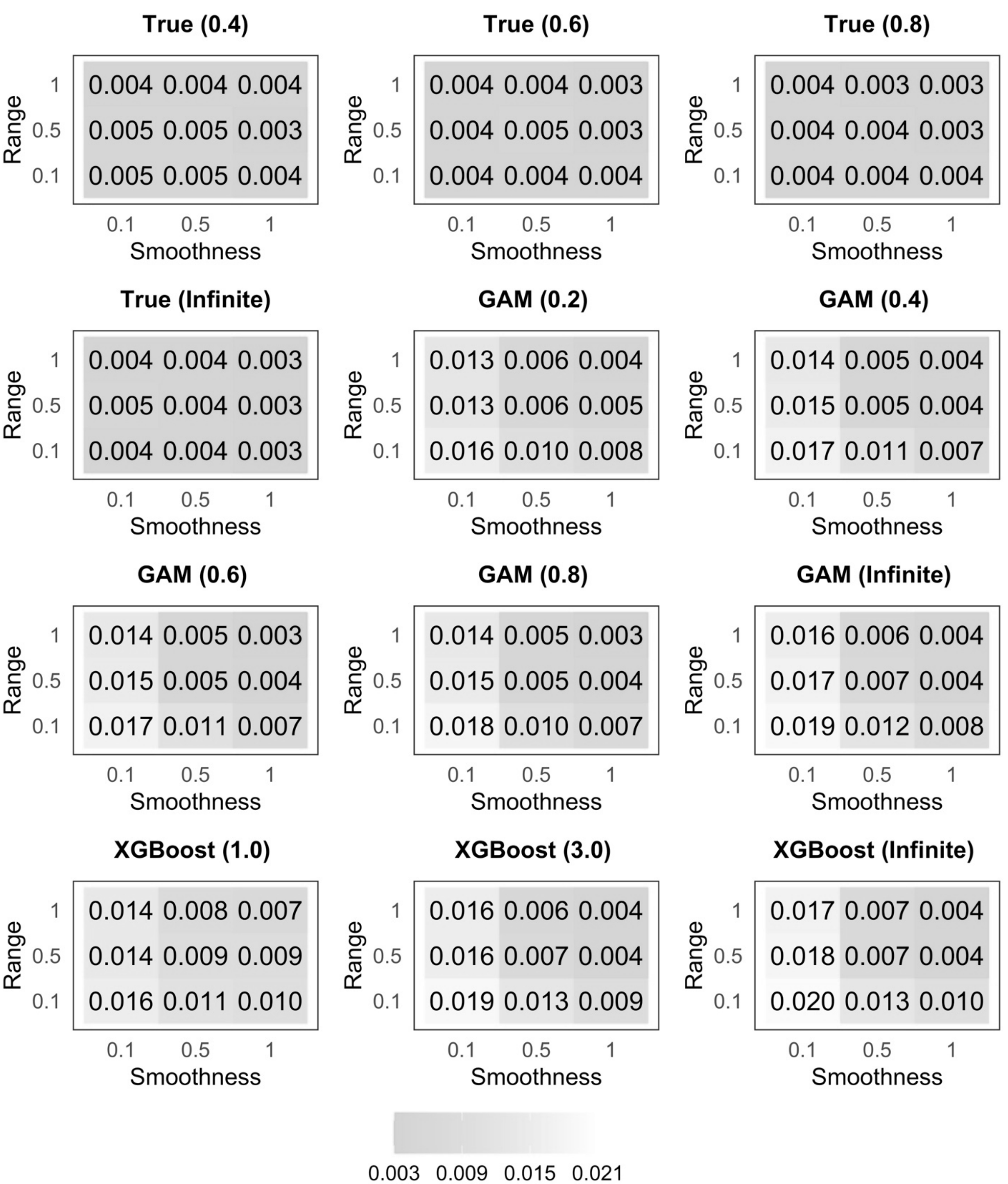

**Figure 7. Root mean squared error by CGPSsm methods with replacement.**

Note: True=True PS and true GPS models in CGPSsm; GAM=GAM with spatial smoother for PS and CGPS estimations; XGBoost=XGBoost for PS estimation and GAM with spatial

smoother for CGPS estimation; The number in brackets indicate $cw$ in one-to-one nearest neighbor caliper matching; $cw = \infty$ indicates one-to-one nearest neighbor matching.

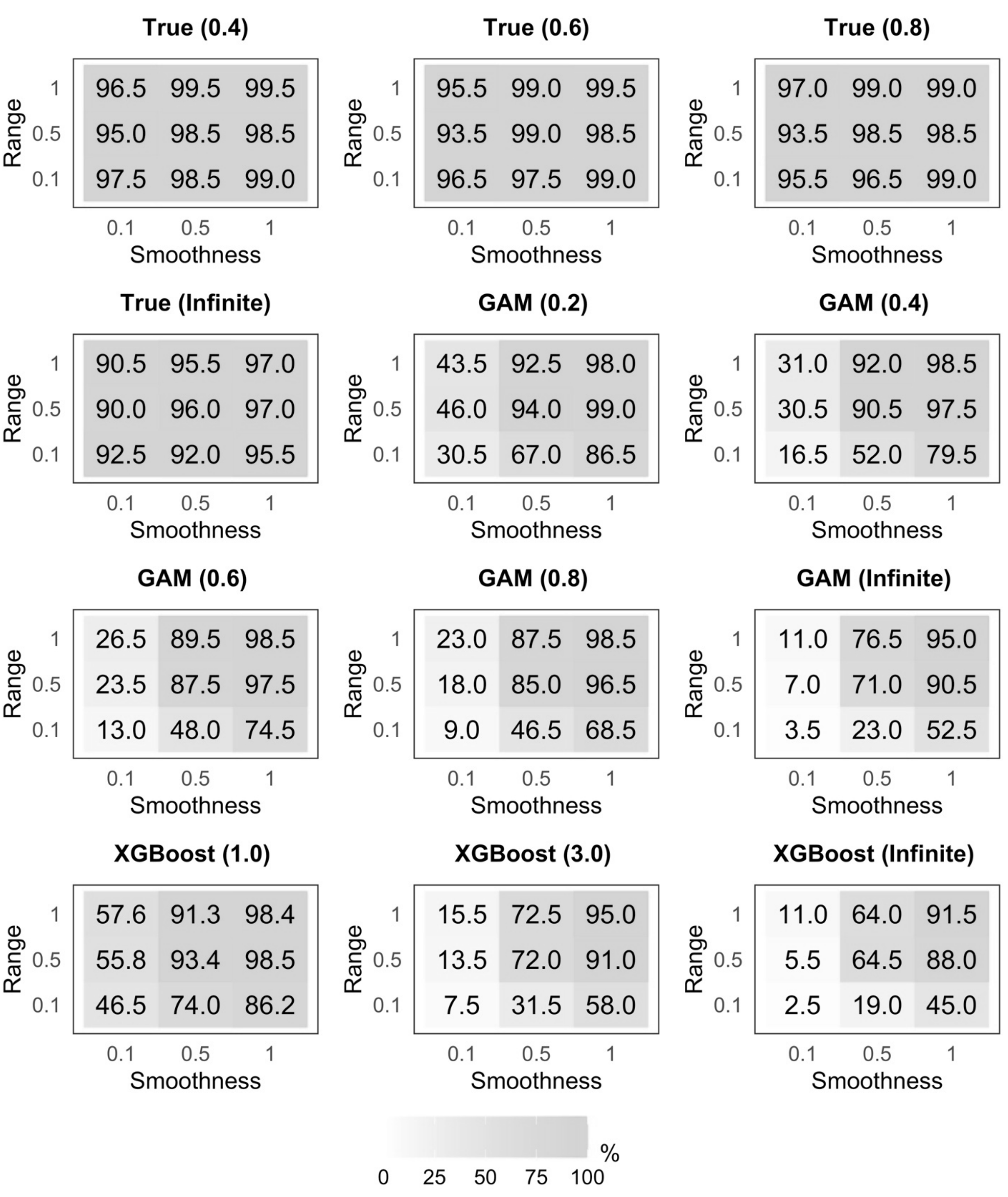

**Figure 8. Nominal coverage of 95% confidence intervals by CGPSsm methods with replacement.**

Note: True=True PS and true CGPS models in CGPSsm (i.e., $U$ is known); GAM=GAM with spatial smoother for PS and CGPS estimations; XGBoost=XGBoost for PS estimation and GAM with spatial smoother for CGPS estimation; The number in brackets indicate $cw$ in one-to-one nearest neighbor caliper matching; $cw = \infty$ indicates greedy one-to-one nearest neighbor matching.

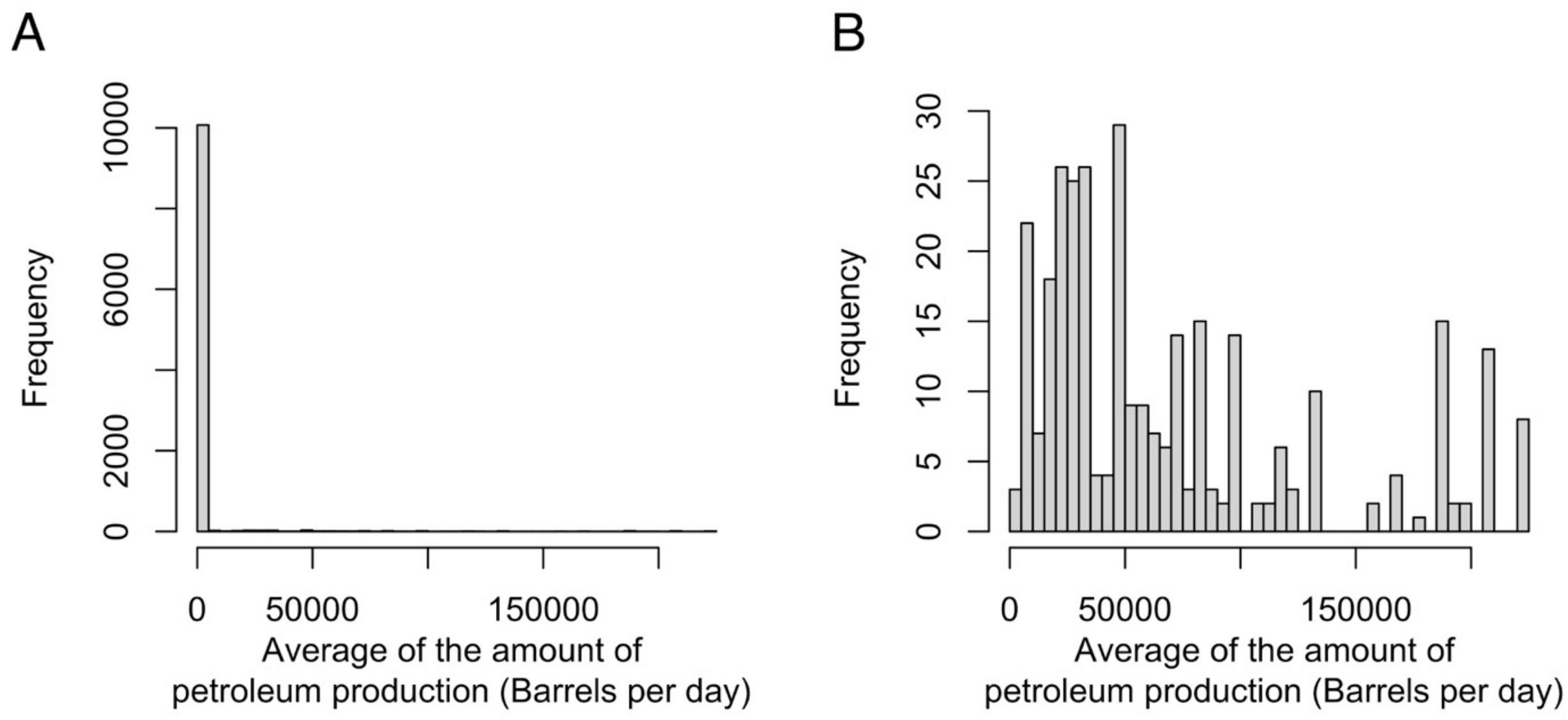

**Figure 9. Exposure distribution in the motivating example: All 10381 census tracts (A); 316 exposed census tracts (B).**

**Supplemental Digital Content**

# Adjustment for Unmeasured Spatial Confounding in Settings of Continuous Exposure Conditional on the Binary Exposure Status: Conditional Generalized Propensity Score-based Spatial Matching

Honghyok Kim, Michelle Bell

School of the Environment, Yale University, New Haven, CT, the United States

**eAppendix 1. Detailed Methods in the Simulation Study**

**Data generating process**

For every pair of two spatial parameters (smoothness parameter, $k$; range parameter, $\pi$) we simulated 200 datasets for which a total of 484 (22×22) fixed locations were simulated as follows.

A. One unmeasured confounder, $U$, was generated from a Gaussian Process with Matérn correlation function. $U$ was normalized to have mean 0 and variance 1.

B. Three observed confounders, $X_1$, $X_2$, and $X_3$ were independently generated from the standardized normal distribution (mean=0; variance=1).

C. A binary exposure, $Z_b$, was generated from a logistic model

$$logit\left(P(Z_b)\right) = -3 + 1X_1 + 1.4X_2 + 0.8X_3 + 1.3U$$

This generating model yields approximately 15% of the observational units exposed.

D. A continuous exposure, $Z_c$ was generated from a linear regression model

$$Z_c = 50 + 2X_1 + 4X_2 + 3.5X_3 + 6U + \epsilon \quad \epsilon \sim N(0{,}5^2)$$

E. An outcome, $Y$, was generated from a Poisson model

$$\log(Y) = 0.03Z_c + 0.15X_1 + 0.23X_2 + 0.31X_3 + U$$

**Matching process and estimation of ATT**

A. One-to-n distance-matching: we matched exposed units to unexposed units if their spatial distance is lower than equal to 0.1 degree. This matching was done with replacement.

B. We fit a PS model with $X_1$, $X_2$, and $X_3$, and coordinates. We fit a generalized additive model (GAM) with the logistic function, binomial distribution, and spatial smoother using coordinates (a thin-plate spline) or used the eXtreme gradient boosting algorithm (XGBoost). For XGBoost, the objective function was a logistic regression for binary classification. The evaluation metric was log-loss. To select hyperparameters (e.g., maximum depth of a tree, maximum number of trees, the learning rate), we used grid-search with 10-fold cross-validation. For the methods with *U* known, we fit a logistic regression model including $X_1$, $X_2$, $X_3$, and *U* instead.

C. We fit a CGPS model with $X_1$, $X_2$, and $X_3$, and coordinates using the exposed subset of the dataset. We fit a GAM with the identity function, gaussian distribution, and spatial smoother using coordinates (a thin-plate spline). For the methods with *U* known, we fit a linear regression model including $X_1$, $X_2$, $X_3$, and *U* instead.

D. We estimated GPS as described in the main text.

E. GPS-matching: For every pair of exposed units and unexposed units matched by the distance, we matched exposed units and unexposed units by GPS. We used one-to-one nearest-neighbor matching with/without replacement and one-to-one nearest-neighbor caliper matching with/without replacement.

F. We generated 500 bootstrapping samples from the distance- and GPS-matched dataset.

G. We fit a disease model to estimate ATT using each of the bootstrapped samples. We used *R* package *gnm* to fit conditional Poisson regression models (with eliminate option for the matched strata). The output of this analysis is equivalent to that of Poisson regression analysis with dummy variables for matched strata added.

H. We obtained mean of coefficient estimates and mean of standard error estimates over the bootstrapped samples.

**eAppendix 2. Detailed Methods in the Application Study**

A. One-to-n distance matching: We matched exposed units to unexposed units if their spatial distance is lower than equal to 0.1 degree. This matching was done with replacement.

B. We fit a PS model with pre-selected measured potential confounders (Table S1), and coordinates. Selection of confounder variables was made considering standardized mean differences. We fit a generalized additive model (GAM) with the logistic function, binomial distribution, and spatial smoother using coordinates (a thin-plate spline) or used the eXtreme gradient boosting algorithm (XGBoost). For XGBoost, the objective function was a logistic regression for binary classification. The evaluation metric was log-loss. To select hyperparameters (e.g., maximum depth of a tree, maximum number of trees, the learning rate), we used grid-search with 10-fold cross-validation.

C. We fit a CGPS model with pre-selected measured potential confounders and coordinates using the exposed subset of the dataset. For exposed units, $X_{ct}$ was log-transformed. Selection of confounder variables was made considering correlations between measured potential confounders and $X_{ct}$ in the exposed subset. We used GAM with the identity function, gaussian distribution, and spatial smoother using coordinates (a thin-plate spline).

D. We estimated GPS as described in the main text.

E. GPS-matching: For every pair of exposed units and unexposed units matched by the distance, we matched exposed units and unexposed units by GPS. We used one-to-one nearest-neighbor matching with/without replacement and one-to-one nearest-neighbor caliper matching with/without replacement.

F. We checked balance in measured potential confounders using standardized mean differences.

G. If covariates were imbalanced, we repeated steps B-F by adding different confounder variables to PS and CGPS models. When they were balanced, we generated 500 bootstrapping samples using the distance- and GPS-matched dataset.

H. We fit a disease model to estimate ATT using each of the bootstrapped samples. We used *R* package *gnm* to fit conditional Poisson regression models (with eliminate option for the matched strata). The output of this analysis is equivalent to that of Poisson regression analysis with dummy variables for matched strata added.

I. We obtained mean of coefficient estimates and mean of standard error estimates over the bootstrapped samples.

**eTable 2.** Data sources and variables used in the application study

| Type | Variable or data | Source |
| --- | --- | --- |
| Outcome | Census tract-level stroke prevalence (%) and coronary heart disease prevalence for the year 2018 | The Centers for Disease Control and Prevention (CDC) PLACES dataset |
| Exposure | Geocoded addresses for 59 petroleum refineries and petroleum production capacity for the year 2015–2017 | The United States Energy Information Administration |
| Confounder (Age) | the percentage of the population aged 18–19, 20–24, 25–44, 45–64, 65–84, and 85 years or older in the population aged 18 years or older | 5-year estimates from the American Community Survey (ACS, 2014–2018) |
| Confounder (Sex) | the percentage of males and females in the population aged 18 years or older | 5-year estimates from the American Community Survey (ACS, 2014–2018) |
| Confounder (Race/ethnicity) | The percentage of Hispanic, non-Hispanic white, non-Hispanic Black. | 5-year estimates from the American Community Survey (ACS, 2014–2018) |
| Confounder (SES) | Median household income, the percentages of the population who live under the federal poverty line or whose highest educational attainment is less than a high-school diploma | 5-year estimates from the American Community Survey (ACS, 2014–2018) |
| Confounder (smoking) | Self-reported current smoking for the year 2018 (%) | The Centers for Disease Control and Prevention (CDC) PLACES dataset |

**eFigure 10. An example of distance-matched datasets with estimated GPS. See Note.**

Suppose that one-to-n distance-matching produced this dataset

| ID | $Z^b$ | $Z^c = w$ | Distance-matched stratum # | $Z^b_{cf}$ | $Z^c_{cf}$ | $GPS \equiv f\left(Z^c_{cf} = w \middle| \boldsymbol{C}\right)$ | Duplicated after the distance-matching with replacement? |
|---|---|---|---|---|---|---|---|
| 1 | 1 | 100 | 1 | 1 | 100 | $f(Z^c = 100 \vert \boldsymbol{C}) =$<br>$P(Z^b = 1 \vert \boldsymbol{C}) \times f(Z^c = 100 \vert \boldsymbol{C}, Z^b = 1) = 0.45$ | No |
| 2 | 0 | 0 | 1 | 1 | 100 | $f\left(Z^c_{cf} = 100 \vert \boldsymbol{C}\right) =$<br>$P\left(Z^b_{cf} = 1 \vert \boldsymbol{C}\right) \times f\left(Z^c_{cf} = 100 \vert \boldsymbol{C}, Z^b_{cf} = 1\right) = 0.20$ | No |
| 3 | 0 | 0 | 1 | 1 | 100 | $f\left(Z^c_{cf} = 100 \vert \boldsymbol{C}\right) =$<br>$P\left(Z^b_{cf} = 1 \vert \boldsymbol{C}\right) \times f\left(Z^c_{cf} = 100 \vert \boldsymbol{C}, Z^b_{cf} = 1\right) = 0.41$ | Yes |
| 4 | 1 | 50 | 2 | 1 | 50 | $f(Z^c = 50 \vert \boldsymbol{C}) =$<br>$P(Z^b = 1 \vert \boldsymbol{C}) \times f(Z^c = 50 \vert \boldsymbol{C}, Z^b = 1) = 0.34$ | No |
| 3 | 0 | 0 | 2 | 1 | 50 | $f\left(Z^c_{cf} = 50 \vert \boldsymbol{C}\right) =$<br>$P\left(Z^b_{cf} = 1 \vert \boldsymbol{C}\right) \times f\left(Z^c_{cf} = 50 \vert \boldsymbol{C}, Z^b_{cf} = 1\right) = 0.33$ | Yes |
| 5 | 0 | 0 | 2 | 1 | 50 | $f\left(Z^c_{cf} = 50 \vert \boldsymbol{C}\right) =$<br>$P\left(Z^b_{cf} = 1 \vert \boldsymbol{C}\right) \times f\left(Z^c_{cf} = 50 \vert \boldsymbol{C}, Z^b_{cf} = 1\right) = 0.32$ | No |
| | | | | | | | |
| … | … | … | … | | | … | … |

After the distance-matching, 1) one-to-one nearest neighbor matching *without* replacement by GPS will produce…

| ID | $Z^b$ | $Z^c = w$ | Distance-matched stratum # | $Z^b_{cf}$ | $Z^c_{cf}$ | $GPS \equiv f\left(Z^c_{cf} = w \middle| \boldsymbol{C}\right)$ |
|---|---|---|---|---|---|---|
| 1 | 1 | 100 | 1 | 1 | 100 | $f(Z^c = 100 \vert \boldsymbol{C}) =$<br>$P(Z^b = 1 \vert \boldsymbol{C}) \times f(Z^c = 100 \vert \boldsymbol{C}, Z^b = 1) = 0.45$ |
| 3 | 0 | 0 | 1 | 1 | 100 | $f\left(Z^c_{cf} = 100 \vert \boldsymbol{C}\right) =$<br>$P\left(Z^b_{cf} = 1 \vert \boldsymbol{C}\right) \times f\left(Z^c_{cf} = 100 \vert \boldsymbol{C}, Z^b_{cf} = 1\right) = 0.41$ |
| 4 | 1 | 50 | 2 | 1 | 50 | $f(Z^c = 50 \vert \boldsymbol{C}) =$<br>$P(Z^b = 1 \vert \boldsymbol{C}) \times f(Z^c = 50 \vert \boldsymbol{C}, Z^b = 1) = 0.34$ |
| 5 | 0 | 0 | 2 | 1 | 50 | $f\left(Z^c_{cf} = 50 \vert \boldsymbol{C}\right) =$<br>$P\left(Z^b_{cf} = 1 \vert \boldsymbol{C}\right) \times f\left(Z^c_{cf} = 50 \vert \boldsymbol{C}, Z^b_{cf} = 1\right) = 0.32$ |
| | | | | | | |
| … | … | … | … | | | … |

Or,

2) one-to-one nearest neighbor matching *with* replacement by GPS will produce…

| ID | $Z^b$ | $Z^c = w$ | Distance-matched stratum # | $Z^b_{cf}$ | $Z^c_{cf}$ | $GPS \equiv f\left(Z^c_{cf} = w \middle| \boldsymbol{C}\right)$ | Duplicated after the distance-matching with replacement? |
|---|---|---|---|---|---|---|---|
| 1 | 1 | 100 | 1 | 1 | 100 | $f(Z^c = 100 \vert \boldsymbol{C}) =$<br>$P(Z^b = 1 \vert \boldsymbol{C}) \times f(Z^c = 100 \vert \boldsymbol{C}, Z^b = 1) = 0.45$ | No |
| 3 | 0 | 0 | 1 | 1 | 100 | $f\left(Z^c_{cf} = 100 \vert \boldsymbol{C}\right) =$<br>$P\left(Z^b_{cf} = 1 \vert \boldsymbol{C}\right) \times f\left(Z^c_{cf} = 100 \vert \boldsymbol{C}, Z^b_{cf} = 1\right) = 0.41$ | Yes |
| 4 | 1 | 50 | 2 | 1 | 50 | $f(Z^c = 50 \vert \boldsymbol{C}) =$ | No |
| | | | | | | $P(Z^b = 1 \mid \boldsymbol{C}) \times f(Z^c = 50 \mid \boldsymbol{C}, Z^b = 1) = 0.34$ | |
| 3 | 0 | 0 | 2 | 1 | 50 | $f\left(Z^c_{cf} = 50 \mid \boldsymbol{C}\right) =$<br>$P\left(Z^b_{cf} = 1 \mid \boldsymbol{C}\right) \times f\left(Z^c_{cf} = 50 \mid \boldsymbol{C}, Z^b_{cf} = 1\right) = 0.33$ | Yes |
| | | | | | | | |
| … | … | … | … | | | … | … |

Note: ID 3 (unexposed unit; in red) is matched to ID 1 (exposed unit; in blue) and to ID 4 (exposed unit; in blue) by one-to-n distance-matching with replacement. GPS of ID 3 in the distance-matched stratum #1 is different than GPS of ID 3 in the distance-matched stratum #2 because $Z^c_{cf}$ is different per $w$ in each stratum. In the GPS-matching step, one-to-one nearest neighbor matching without replacement will give us a pair of ID 1 and ID 3 at the distance-matched stratum #1 and will give us a pair of ID 4 and ID 5 at the distance-matched stratum #2. For the latter, although GPS of ID 3 (0.33) is closer to GPS of ID 4 (0.34) than GPS of ID 5 (0.32), ID 4 is matched to ID 5 because ID 3 is already matched to ID 1 at the distance-matched stratum #1. In contrast, one-to-one nearest neighbor matching with replacement will match ID 4 to ID 3.

**eFigure 11. Spatial patterns of *U* with nine pairs of the smoothness and range parameters of the Matérn covariance function.**

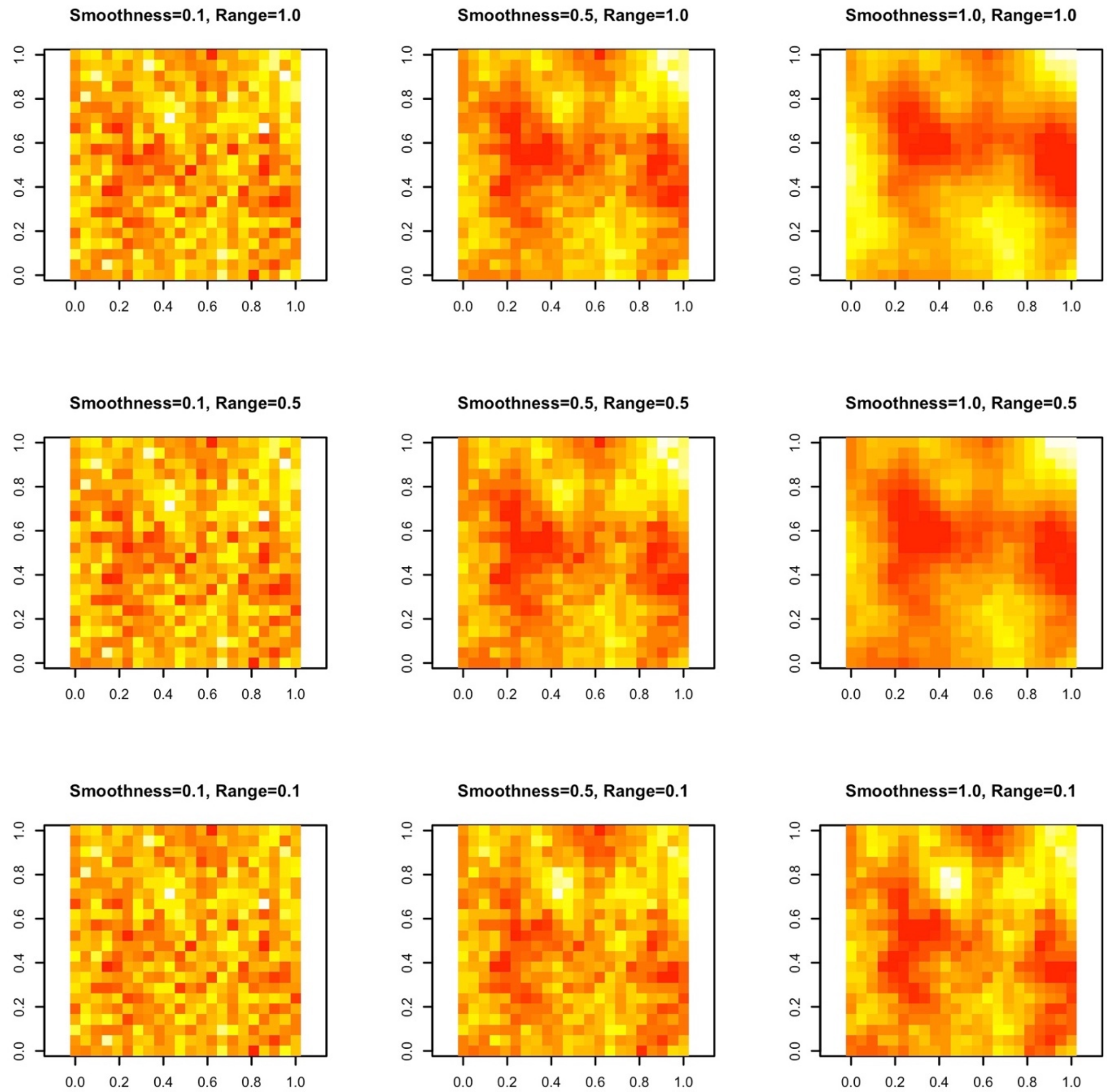

**eFigure 12. Bias by traditional regression with/without *U*, Naïve IPW-GAM, and Naïve IPW-XGBoost.**

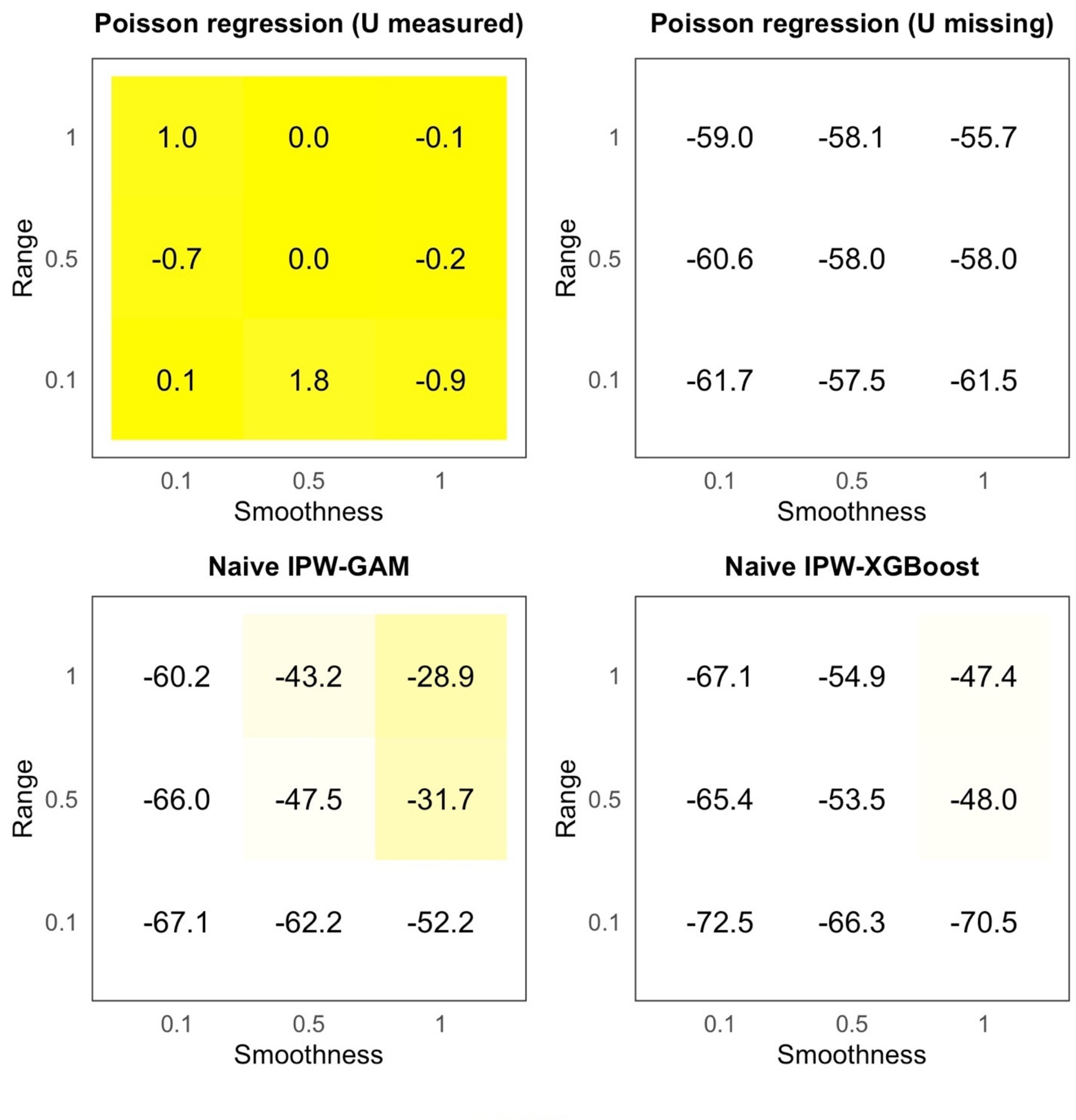

**eFigure 13. Nominal coverage of 95% confidence intervals by traditional regression with/without *U*, Naïve IPW-GAM, and Naïve IPW-XGBoost.**

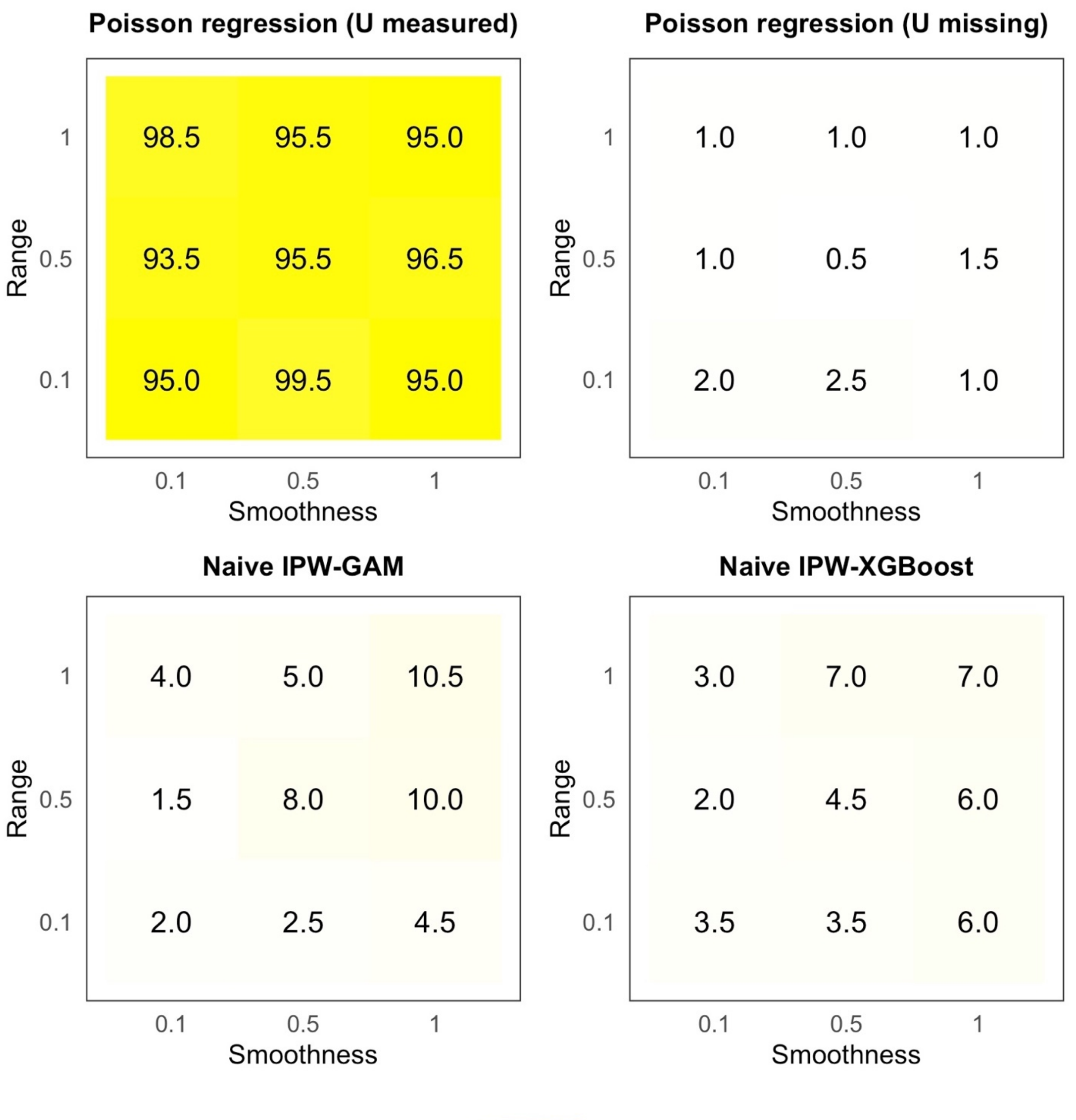

**eFigure 14. Bias by CGPSsm methods without replacement.**

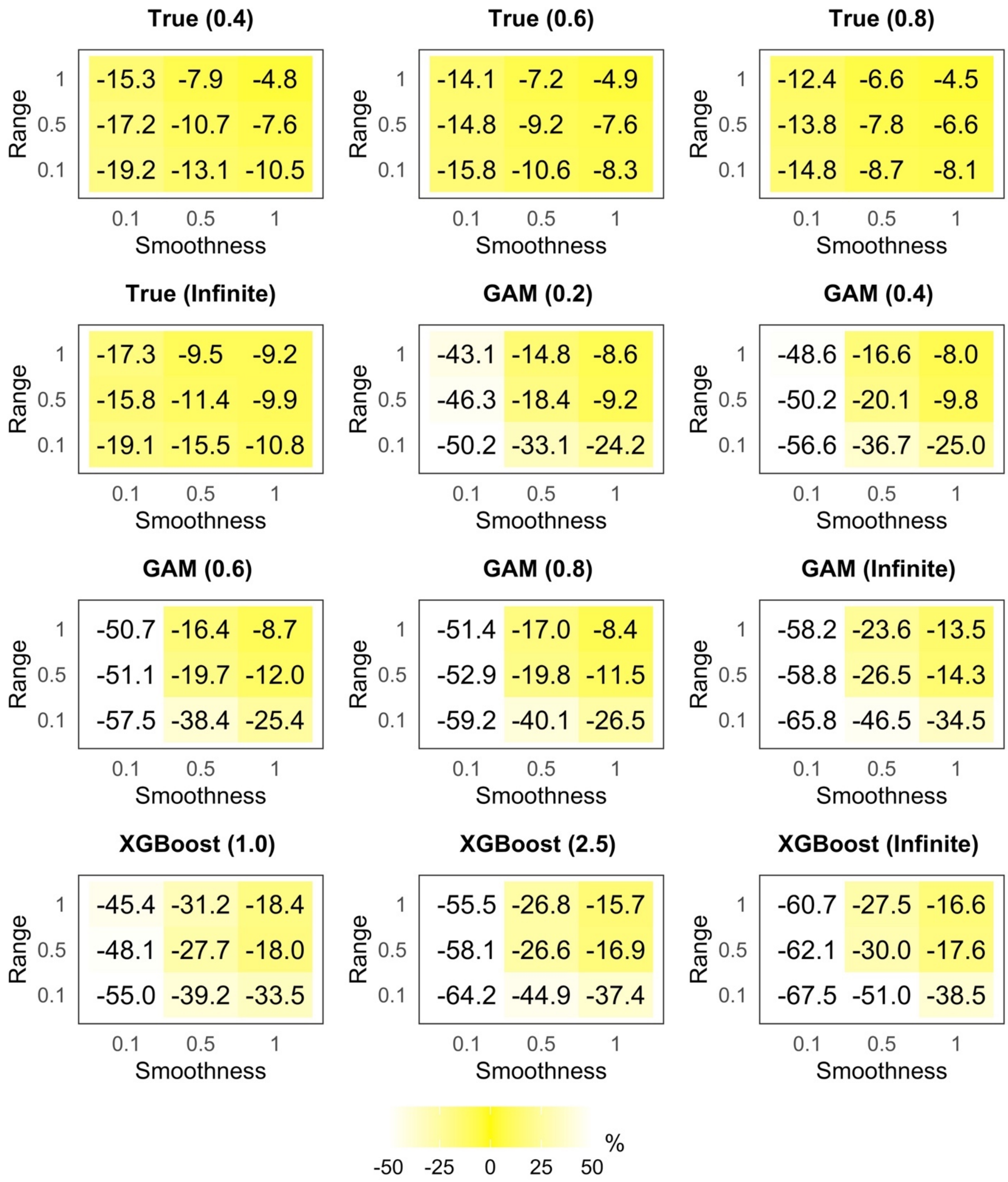

**eFigure 15. Root mean squared error by CGPSsm methods without replacement**

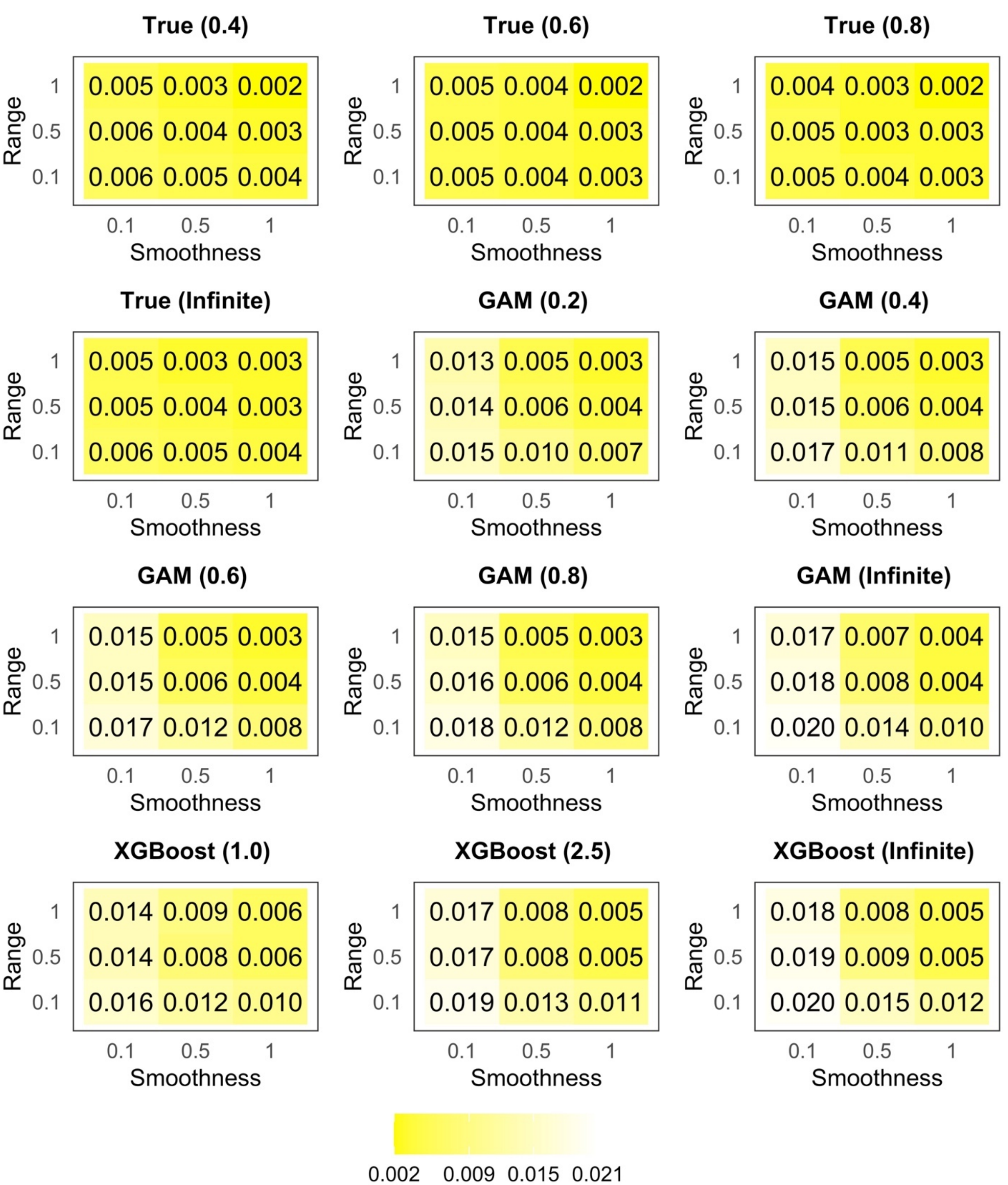

**eFigure 16. Nominal coverage of 95% confidence intervals by CGPSsm methods without replacement.**

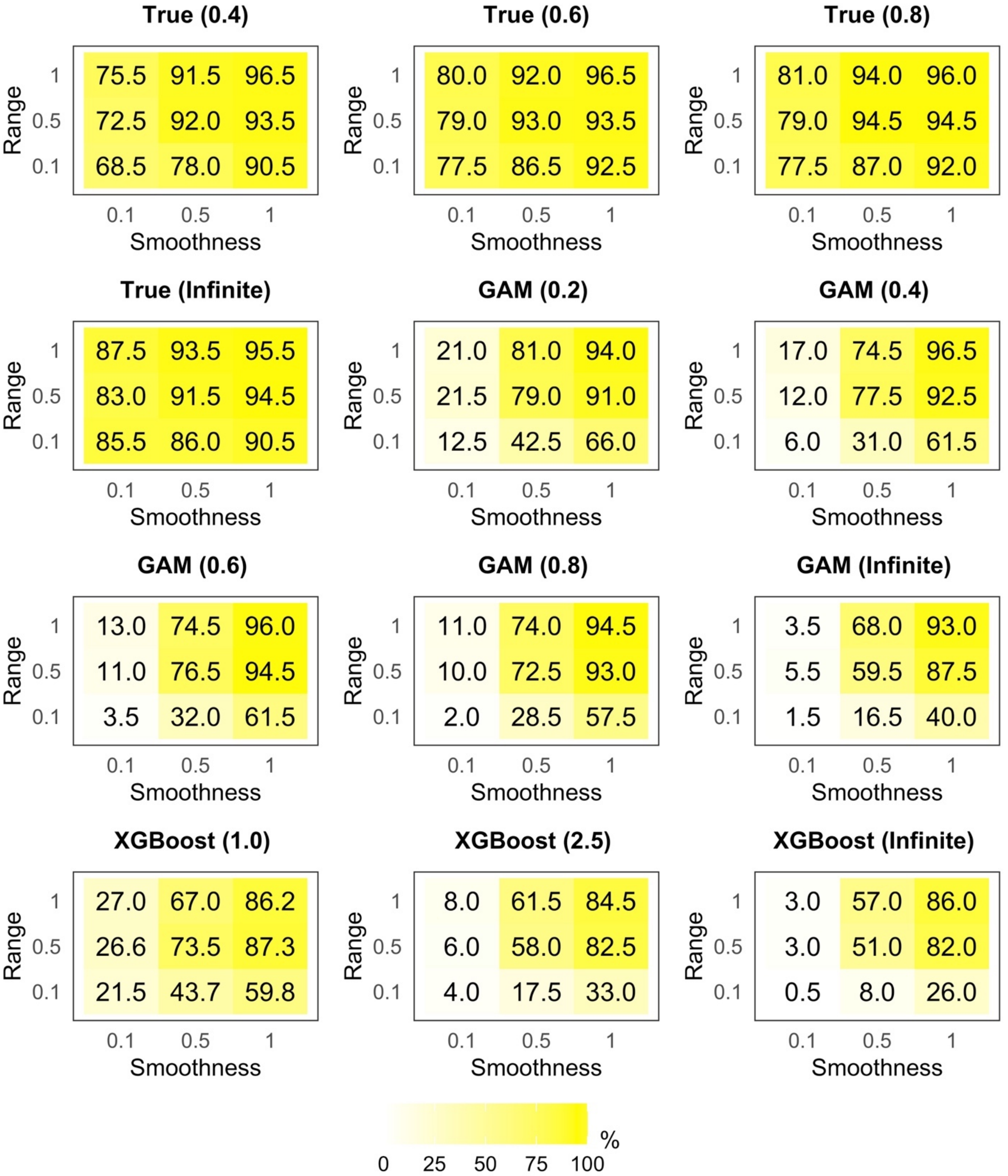

**eFigure 17. Percentage of exposed units matched to unexposed units by CGPSsm methods without replacement.**

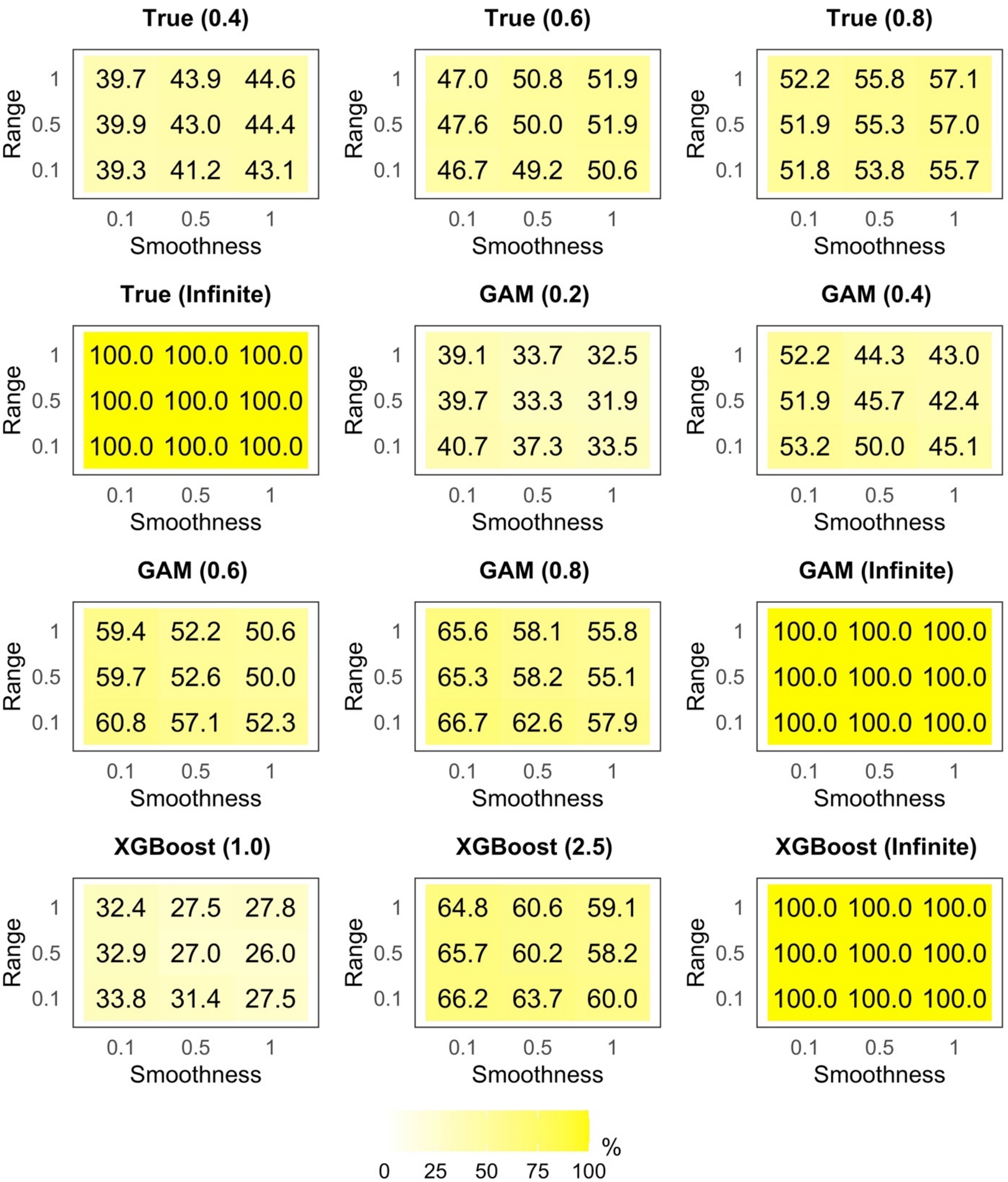

**eFigure 18. Covariate balance before and after CGPSsm without replacement.**

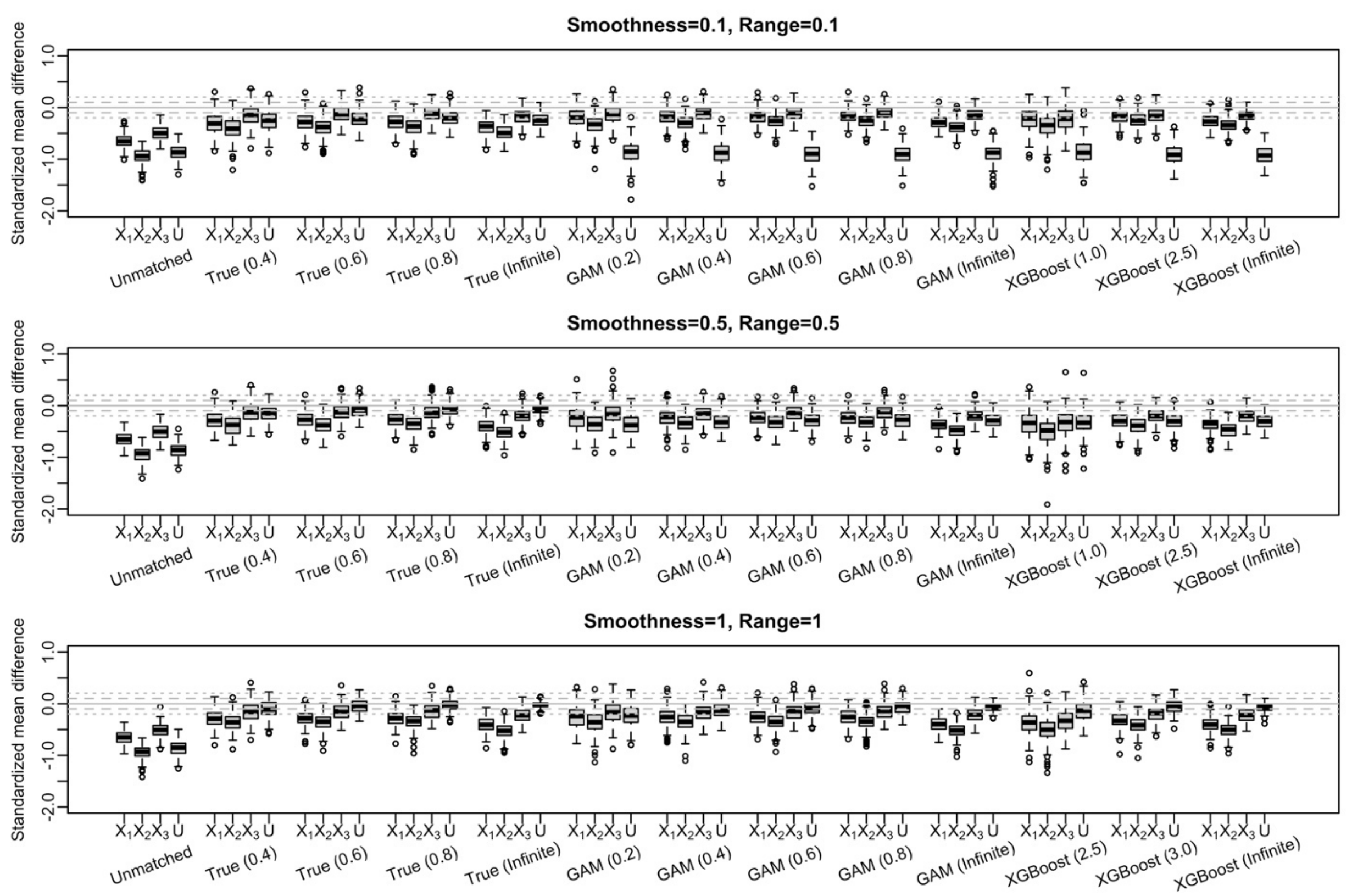

**eFigure 19. Covariate balance before/after CGPSsm in the motivating example.**

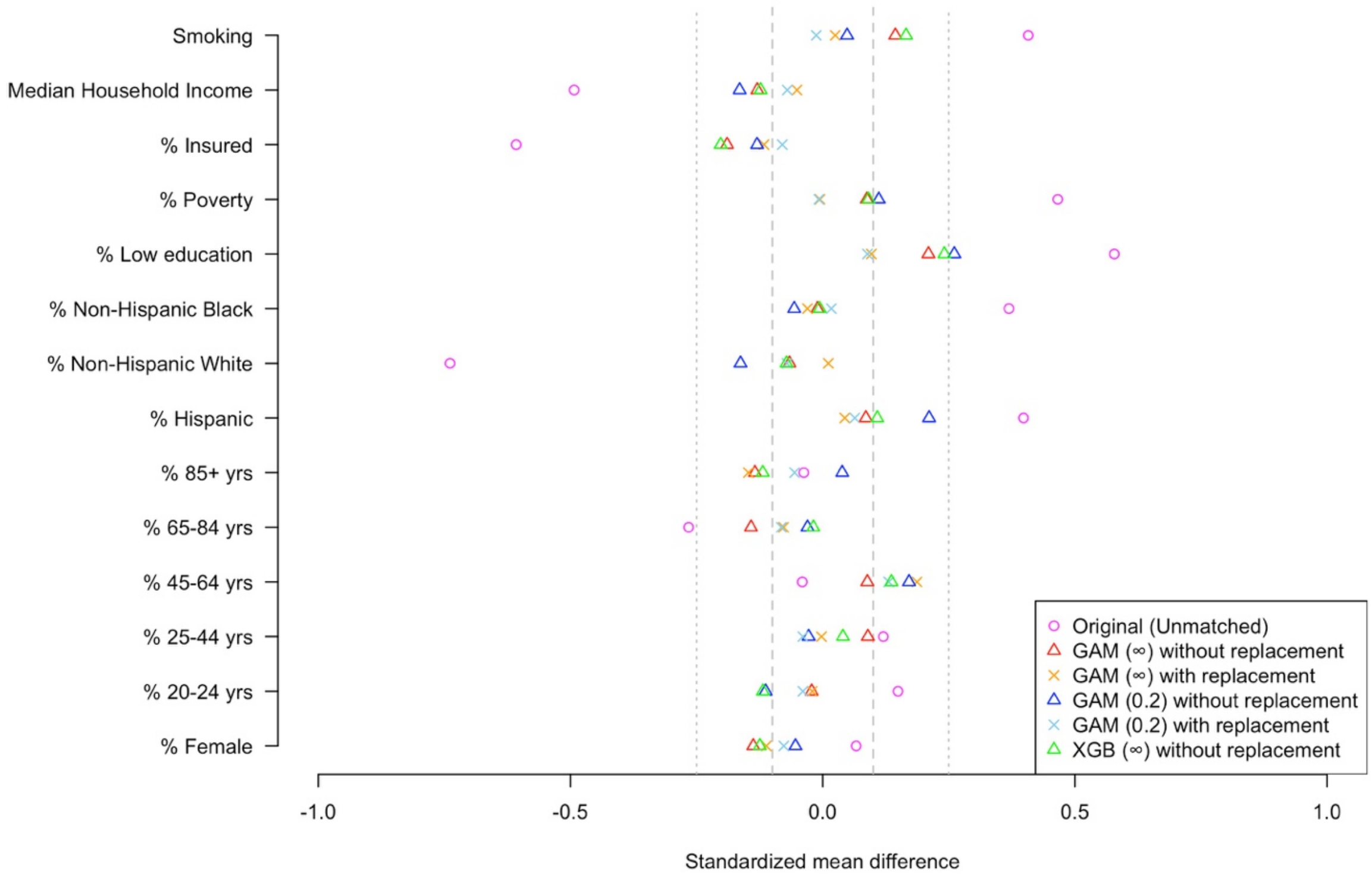

Note: The number in brackets indicate $cw$ in one-to-one nearest neighbor caliper matching; $cw = \infty$ indicates one-to-one nearest neighbor matching; Grey dashed lines indicate ±0.1. Grey dotted lines indicate ±0.25.